\documentclass[letterpaper,aps,preprintnumbers,longbibliography,notitlepage,longbibliography]{revtex4-1}
\usepackage[latin9]{inputenc}
\usepackage{xcolor}
\usepackage{amsmath}
\usepackage{amssymb}
\usepackage{graphicx}
\usepackage{esint}
\usepackage{stmaryrd}
\usepackage{subfigure,epsfig,amsfonts}
\usepackage{natbib}
\usepackage{amsthm}
\usepackage{dsfont}
\usepackage{braket}
\usepackage{hyperref}

\makeatletter

\pdfpageheight\paperheight
\pdfpagewidth\paperwidth

\pdfoutput=1

\usepackage{epsfig}
\usepackage{graphics}

\usepackage{mciteplus}
\mciteErrorOnUnknownfalse

\makeatother
\graphicspath{ {figures/} }

\begin{document}

\preprint{FERMILAB-PUB-16-264-A}

\title{Statistical Model of Exotic Rotational Correlations in Emergent Space-Time}

\author{Craig  Hogan}
\affiliation{University of Chicago and Fermilab}
\author{Ohkyung Kwon\footnote{E-mail: o.kwon@kaist.ac.kr}}
\affiliation{Korea Advanced Institute of Science and
Technology}
\author{Jonathan Richardson}
\affiliation{University of Chicago and University of Michigan}


\begin{abstract}
A Lorentz invariant statistical model is presented for rotational fluctuations in the local inertial frame  that arise from new quantum degrees of freedom of space-time.
The model assumes invariant classical causal structure, and a Planck information density in  invariant proper time determined by the world line of an observer.   It describes macroscopic spacelike correlations  that appear as observable timelike correlations in  phase differences of light propagating on paths that begin and end on the same world line. The model  allows  an exact prediction for the   autocorrelation of any interferometer time signal  from the shape of the light paths.  Specific  examples  computed for configurations that approximate realistic experiments show that the model can be rigorously tested, allowing a direct experimental probe of Planck scale degrees of freedom.
 \end{abstract}
\maketitle

\section{Introduction}

In both classical special relativity and quantum mechanics,   a physical  rotation, or rate of change of direction, has a definite physical meaning  even on infinitesimal scales.
The  ``local inertial frame'' is an absolute and determinate physical  attribute of space, measurable for example by an absence of centrifugal acceleration.
However, when quantum mechanics is combined with general relativity, the use of a classical local inertial frame as a reference to define rotation becomes inconsistent, even in a flat space-time background. Because of gravitational frame dragging, space itself inherits the quantum indeterminacy of any apparatus used to measure rotation, an effect that becomes more pronounced on short time and length scales.   Extrapolation  of standard quantum principles\cite{Hogan:2015sra} suggests that     rotational observables of space-time are quantized at the Planck length $l_P\equiv \sqrt{\hbar G/ c^3}= 1.616\times 10^{-35}$ meters, where  a single quantum has the energy and angular momentum of a maximally rotating black hole.
As in any quantum system, the  quantization of space-time rotation should manifest  as new forms of  correlations in rotational observables.
We call these ``exotic'' rotational correlations because they are not part of a standard quantum system of particles or fields, but arise from  quantum degrees of freedom that give rise to space and time. If they 
could be measured, they would provide experimental clues to the nature of those  degrees of freedom.

In this paper, we develop  a statistical model to predict the properties of exotic rotational  correlations in  systems much larger than the Planck length. 
A purely rotational structure of  correlations in space and time is fixed by the Lorentz invariant classical causal structure,
with a normalization fixed by a Planck scale information density in invariant proper time.
The finite information in the system leads to  exotic fluctuations  in measurements of rotation  based on the propagation of light. In an interferometer,   exotic spacelike correlations lead to exact predictions for 
 timelike correlations of  signals that  depend only  on the spatial
 layout of the light paths in space and  the  Planck  scale information density. The model thus presents a definite, quantitative  hypothesis  for  how inertial frames,  rotation and directionality emerge statistically  from quantum geometry.

 Our model is embedded in a flat, classically stationary  space-time, but  its correlations  on large scales  match  the Planck scale holographic information density on  causal diamond surfaces required for emergent theories of gravity\cite{Jacobson1995,Verlinde2011,Padmanabhan:2013nxa}.
It is natural to  conjecture  that  the specific exotic rotational degrees of freedom identified in our model describe  the ground state of the quantum system that, when excited, gives rise to  fully dynamical  quantum gravity\cite{Rovelli2004,Thiemann:2007zz,Ashtekar:2012np}. If this conjecture is correct, it may be possible to study some symmetries of quantum gravity with precision laboratory measurements.

 Our analysis will be useful to design and interpret experiments based on correlations in interferometer signals, which now attain the sensitivity needed\cite{Holo:Instrument} to measure the predicted exotic correlation.  As in  previous  estimates of holographic geometrical fluctuations with similar displacement magnitude\cite{Hogan2008a,Hogan2008,Hogan2012,2013ASPC..467...17H,Kwon:2014yea,Hogan:2015kva}, 
directions in this model on scale $R$ fluctuate on timescale $\approx R/c$,  with a variance of about  $\langle \Delta\theta^2\rangle_R\approx l_P/R$, which leads to a signal with  a  power spectral density of fractional displacement noise about equal to a Planck time\cite{Holo:Instrument}.   
However,  purely rotational fluctuations would not have been detected in  current experiments that constrain holographic fluctuations with this sensitivity\cite{Holo:PRL}, due to their purely radial light paths.  As shown below, a  definitive signature of rotational correlations can be measured using modified experimental configurations.  

Our model provides a concrete, quantitative example of how Planck scale departures from a classical space-time might produce detectable physical effects on much larger scales, due to large scale correlations implied by holographic information bounds. This specific model can be definitively tested with experiments, but it may not be unique: the effects of geometrical correlations on quantum states of light and matter in an apparatus are not all rigorously constrained by known basic principles. Ultimately, the best hope is that the symmetries of Planck scale geometry and their relationships to the quantum states of Standard Model fields can be probed by a systematic experimental program involving a variety of instruments, including other forms of interferometry capable of measuring superluminally time-resolved correlations with Planck sensitivity. If exotic rotational geometrical fluctuations in any form are shown to exist in laboratory experiments, they are also likely to play an important role in shaping other physical systems, such as the cosmological constant in an emergent metric with a matter field vacuum\cite{Hogan:2015sra}, or the generation of large scale metric fluctuations during an inflationary period in the early universe. Such effects depend on entanglement of the geometrical system with field vacuum states, and the back reaction on the emergent metric, which lie beyond the scope of the model developed here.

\section{Exotic Rotational Correlations}

\subsection{Statistical Lorentz Invariance and Observables}

A widely known difficulty in quantizing gravity is preserving local Lorentz invariance. Our model is motivated by quantum gravitational considerations, but we do not quantize gravity here, so our task is considerably easier. 
The model is constructed in a special relativistic space-time, in the limit of a fixed causal structure with no curvature or dynamics. 
The correlations are defined by spacelike displacements on light cones, which are invariant structures.  Our calculated examples assume a nonaccelerating observer for realism and convenience, but since the  framework depends only on invariant causal structure,  it  generalizes to any timelike observer. 

Thus, the model preserves the principle of Lorentz invariance in a statistical sense\cite{Hogan:2015sra}. Quantum mechanics can only make predictions about correlations, that is, statistical outcomes of  measurements. We abandon the classical notion of an observer-independent, determinate and absolute space-time, but require that the correlations predicted by the theory do not depend on arbitrary choices of coordinates or frames. This property is built in by construction: the correlations in our model depend only on invariant classical causal structures. 

The invariant causal structures are anchored to the specific world line of a measurement,  identified as the quantum-mechanical ``observer.'' On this world line, we posit that proper time intervals--- Lorentz scalars, invariantly and locally definable--- are quantized at the Planck scale. By imposing the Planck scale information density in  proper time of a measurement, we avoid the Lorentz invariance violation that occurs from
 a  Planck scale imposed  on  timelike or spacelike intervals that depend on a choice of coordinate system.

Our predictions are given in terms of locally measurable  observables. We present here observable statistics about correlations in proper time and frequency. Any pair of events viewed in any frame (boosted relative to the observer, say) is separated by the same invariant interval, so the predictions are invariant Lorentz scalars. The calculation involves  quantities  that depend on the observer's nonlocal positional relationships with other trajectories, but  begins and ends with Lorentz scalars. 

\subsection{Model assumptions}

 The model is based on  two principal assumptions:
 \begin{enumerate}
 \item 
The system has a  Planck scale information density, in invariant proper time, on the world line of a measurement.

Unlike a classical world line,  the information content or bandwidth of a  world line, and any measurement,  is limited to that of a  discrete 1D time series with steps of  duration $\approx l_P/c$ in invariant proper time.
 That limit applies as well to the rotational relationship of its local inertial frame with rest of the universe. The  information available to describe directional orientation relative to an interval on a world line  matches its duration in Planck units.
 
\item
The quantum departure from a classical space-time can be represented by random Planck scale spacelike displacements that exactly preserve classical causal structure defined by the light cones of the world line of the measurement.
 
 The exotic degrees of freedom are assumed to act physically as random Planck scale transverse position displacements of events.
 The displacements have zero mean fixed by the classical limit, and a nonzero variance from quantum fluctuations. 
They are not independent, but must be highly correlated within a Planck time, because of the Planck information density. Correlations in a measurement are determined by the invariant light cones of a measurement interval.
Because of the limited information in any finite interval, measurements of rotation fluctuate from the absolute classical inertial frame. 
While null relationships between events are exactly preserved, timelike and spacelike relationships depend on preparation and measurement of a state.  
 
\end{enumerate}

These are the only properties of the quantum geometry necessary to specify the exotic statistical effects on propagating light in flat space-time.
The Planck scale normalization is fixed by requiring that the number of degrees of freedom of a causal diamond agrees with holographic gravity. 
The system  approaches a standard, flat space-time on  scales much larger than the Planck length, in a nontrivial  way  that is uniquely fixed by these assumptions. The model  accounts for  how  a nearly classical  inertial frame around any world line  emerges over a long time interval from indeterminate rotation  at the Planck scale.  

\subsection{Covariance of Random Displacement on Future Light Cones of an Observer}

\begin{figure}[ht]
    \centering
    \includegraphics[width=0.78\linewidth]{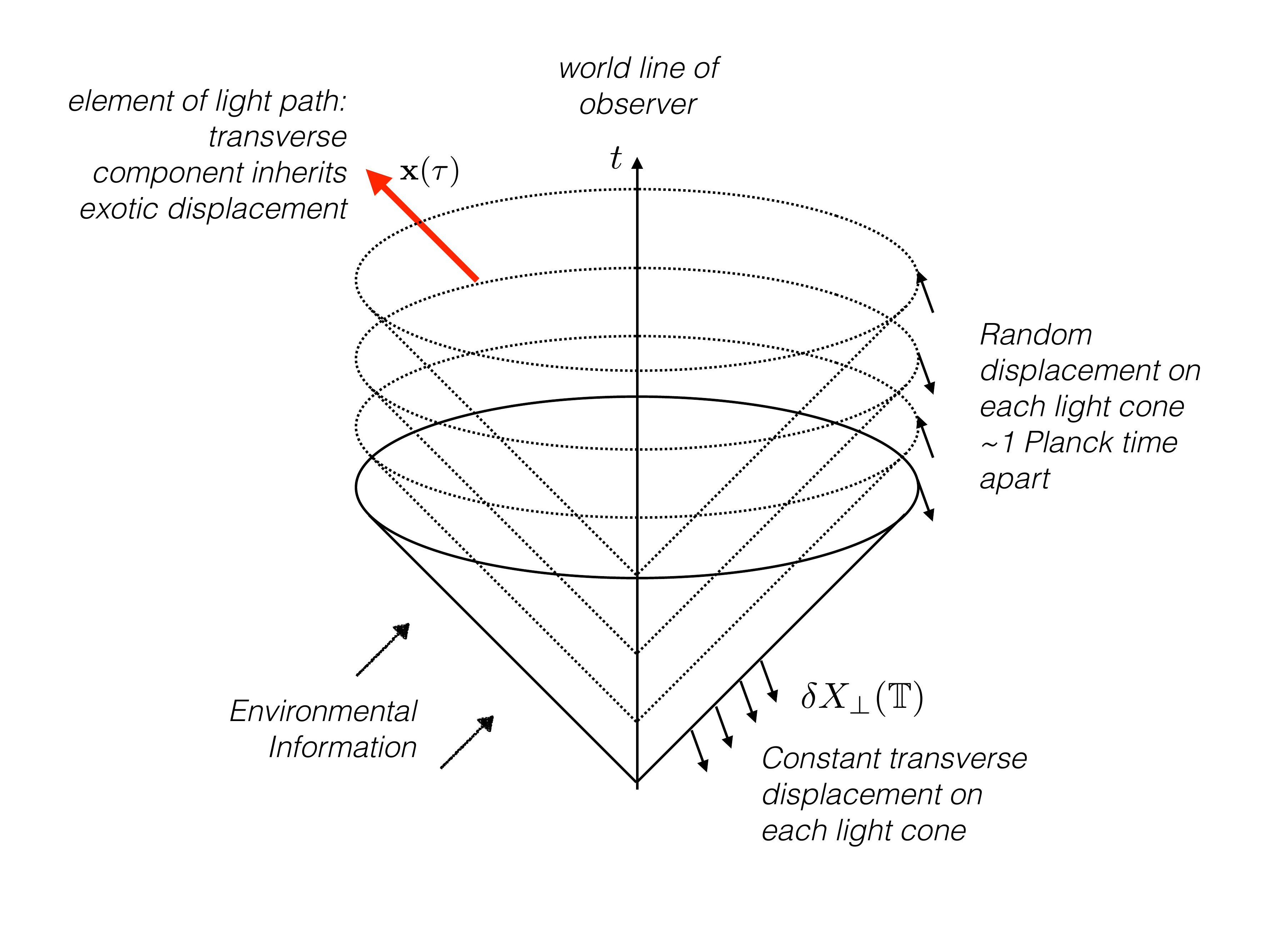} 
    \caption{Classical picture of the causal structure of large scale covariances in exotic space-time positional relationships.  Space and time are shown in the observer's frame, with one spatial dimension suppressed.  A series of light cones, surfaces of constant light cone time $\mathbb{T}$, is shown emerging from an observer's world line. These determine the space-time structure of correlations produced by Planck scale entanglement.  Within a Planck length of each surface of constant $\mathbb{T}$, the same  random transverse spatial displacement or ``quantum twist'' $\delta X_{\perp}(\mathbb{T})$ of about a Planck length--- a consequence of noncommutative geometry at the Planck scale---  is causally inherited by entanglement with a quantum state prepared on the observer's world line. Thus,  entanglement is causally local: each event on the observer's world line determines the projection of all future events that lie within a Planck length of null separation. Propagating light  inherits the projection of the local transverse displacement onto its path $\mathbf{x}(\tau)$.  The total effect on the light phase is an accumulation of projected components of random displacements from crossing a series of light cones.  Transverse displacements accumulate like a random walk and lead to a  mean square displacement much larger than the Planck length.}
    \label{fig:twists}
\end{figure}

\label{sec:cts_coords}

Light cones (or null cones) are the covariant  objects that define causal structure: they define the sharp classical  boundaries between past, present and future, and between timelike and spacelike separation of events. We base our model on the principle that classical causal structure is respected on all scales larger than the Planck length.
 In  the frame of an observer $O$, we define a ``light cone time'' coordinate variable, 
\begin{equation}
\mathbb{T}_O=  t_O-\frac{|\mathbf{x}_O|}{c}.\label{eq:BSlightconetime}
\end{equation}
A surface of constant $\mathbb{T}$ in 3+1D represents a 2+1D light cone emanating from  an event on an observer's world line, a 3-surface defined in conventional coordinates by $t=|\mathbf{x}|/c$.  Although we will choose to calculate in the rest frame of an observer, the causal relationships defined by the actual light cones  are  independent of the choice of frame and have a Lorentz invariant physical meaning.    In the following, we drop subscripts until they are needed later for comparing measurements from more than one observer.

To connect with physical observables we  develop a model of the geometrical character of  random variables, which will allow us to compute how the random displacements affect physical quantities. The exotic  
 departure from the classical system is described as a random variable with  variance $\ell_{P}^2$:
\begin{eqnarray}\label{eqn:variance}
 \delta X_{\bot}(\mathbb{T})\sim\mathcal{N}\left(0,\ell_{P}^{2}\right).
\end{eqnarray}
The transverse displacement variable $\delta X_{\perp}$  represents an exotic  displacement relative to the classical inertial frame, normal to the direction to the observer. It has zero mean so the large scale average is classical.  In our model it is interpreted physically as a displacement of phase for transversely propagating light, but should be regarded as a modification of  classical position of space itself, and everything in it, relative to the observer. The coherence scale and randomness  model the effect of the finite information content of the indeterminate quantum elements. Since the physical effects are purely transverse, the model exactly preserves causal structure,  an assumed exact symmetry of quantum gravity. 


In a discrete description with the same information content, this quantity corresponds to an exotic incremental displacement at each discrete step--- a random step of length $\ell_P$ each  timestep $t_P$, due the noncommutativity of discrete quantum geometrical operators. 
All of the predictions below have only  one parameter, $\ell_P$, which, to be consistent with a discrete description and exact causal symmetry, denotes both
the  
coherence length of the quantum elements and the magnitude of  random displacements. A non-zero value of  $\ell_{P}$ results in a finite, countable information content, and introduces deviations from classical continuous space-time.  Based on extrapolation of gravity and quantum mechanics\cite{Hogan2008a,Hogan2008,Hogan2012,2013ASPC..467...17H,Kwon:2014yea,Hogan:2015kva,Hogan:2015sra}, its value should be
approximately the  Planck length $l_P$.

We now introduce the central  hypothesis  that encapsulates the relationship of Planck scale quantum elements to each other and to an observer on large scales (see Fig. \ref{fig:twists}). 
That hypothesis can be written as a  covariance of an exotic displacement on
future light cones:
\begin{equation}\label{eqn:covariance}
\textrm{cov}\:\Big(\delta X_{\perp}(\mathbb{T}'),\,\delta X_{\perp}(\mathbb{T}'')\Big)=\begin{cases}
\ell_{P}^{2}\;, & \left|\mathbb{T}''-\mathbb{T}'\right|<\frac{1}{2}t_{P}\\
0\;, & \textrm{otherwise}
\end{cases}
\end{equation}
For calculational convenience and clarity, in place of a discrete model of time steps, or a continuous Gaussian roll-off (say), the model uses sharply delineated bins of constant covariance in $\mathbb{T}$; none of the results depend on this choice.

Although we call the displacements ``quantum twists,'' note that  the variable represents a constant transverse displacement, and not a constant angle, everywhere on the light cone, independent of separation from the observer. The  correlations that result from entanglement appear spooky, since they appear nonlocally across spacelike hypersurfaces,  but they are still local in the sense that they are determined by sub-Planck  light cone time separation between events.
Causal relationships determine the collapse of the quantum state into  classical random observables. 

Although directions at the Planck length are indeterminate (and if measured, fluctuate by about a radian per Planck time), nearly classical directions emerge in a large scale average.  Directions to distant objects and events are not independent, but are entangled with each other; those at distance $R$ can be encoded by $\approx (R/\ell_P)$ qubits.  
Information from each event  on a world line prepares the state and  entangles it with the projection of the displacement operators at all events on its light cone.
 Our use of a random variable  is a classical shorthand for quantum decoherence,  and our covariance is a classical shortcut to describe quantum entanglement among the individual Planck scale elements:  events on a light cone  must  ``agree on'' the
projection of the corresponding spin state onto any given axis. In the plane of an interferometer, every point on a light cone steps clockwise or counterclockwise by
one Planck length from the previous one.  Apparently non-local spatial correlations arise from causal entanglement
of the Planck scale degrees of freedom.

\label{sec:fluctuation_rate}

It will prove further convenient to define a \textit{rate} of transverse
position fluctuation, or an effective transverse ``velocity'' or jitter,
\begin{eqnarray}
 \frac{dX_{\perp}(\mathbb{T})}{d\mathbb{T}}\equiv\frac{\delta X_{\perp}(\mathbb{T})}{t_{P}}
\end{eqnarray}
This quantity  is  not a physical velocity or momentum, but  represents a discrete ``motion'' of the Planck scale quantum elements with respect to the classical system.
It is explicitly dependent only on $\mathbb{T}$ (and not individually
on $t$ or $\mathbf{x}$) due to the coherence of the fluctuations
on light cones. This variable is distributed as 
\begin{eqnarray}
 \frac{dX_{\perp}(\mathbb{T})}{d\mathbb{T}}\sim\mathcal{N}\left(0,\,\left(\frac{\ell_{P}}{t_{P}}\right)^{2}\right)
\end{eqnarray}
with the covariance structure 
\begin{equation}
\textrm{cov}\left(\frac{dX_{\perp}}{d\mathbb{T}}(\mathbb{T}'),\,\frac{dX_{\perp}}{d\mathbb{T}}(\mathbb{T}'')\right)=\begin{cases}
{\displaystyle \left(\frac{\ell_{P}}{t_{P}}\right)^{2},} & \left|\mathbb{T}''-\mathbb{T}'\right|<\frac{1}{2}t_{P}\\
0\;, & \textrm{otherwise}
\end{cases}
\end{equation}
The characteristic rate of a transverse position random walk is thus one coherence
length per coherence time, or $c$.

\section{Exotic Correlations in Interferometer Signals}

We  now estimate the effect of exotic displacements
 on measurable phase shifts in beams of freely-propagating
light. These phase shifts are measured by an  interferometer consisting 
of a beamsplitter and two or more reflecting mirrors. Light from a
coherent source is split into two beams which propagate through an
arrangement of mirrors and back to the beamsplitter, where they recombine.
Any difference in the distance propagated by the two beams introduces
a corresponding difference in the accumulated optical phase. Interference
effects from this phase difference modulate the power of the recombined
beam, which is read out via a photodetector at the output, or antisymmetric
port of the instrument. In this section, the general response
of an interferometer to exotic rotational displacements
will  be derived. In spite of the simple and symmetric form of the fundamental exotic covariance, the projection onto observables is surprisingly subtle. These results will be used below to compute the instrument response of specific configurations of mirrors.

\begin{figure}[ht]
    \centering
    \includegraphics[width=0.77\linewidth]{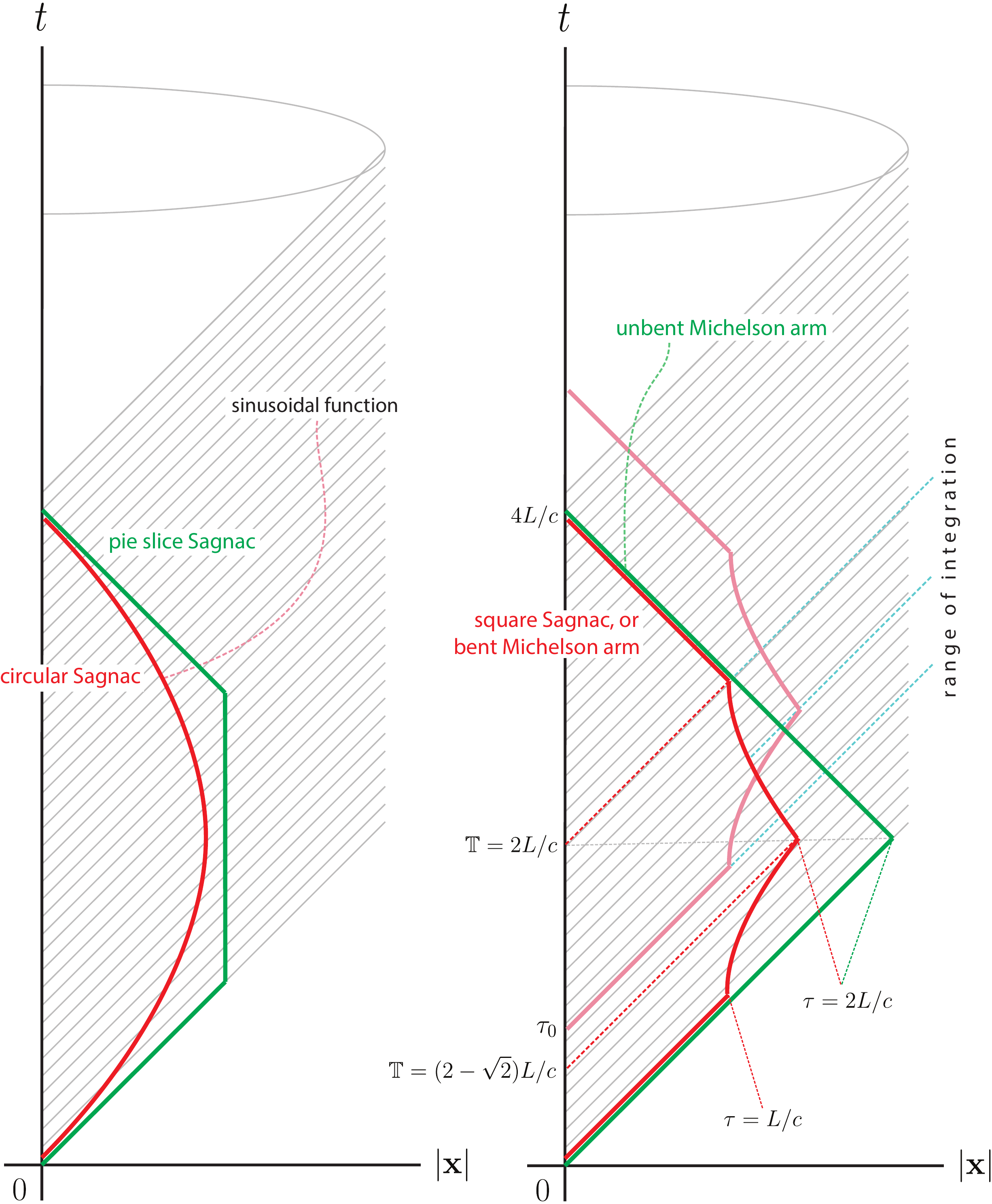}
    \caption{Space-time diagram showing how  quantum twists of the inertial frame project onto an interferometer signal. Laboratory time is plotted against radial position in the observer frame, with the two transverse spatial dimensions suppressed, so each point represents a 2-sphere at a single time. A sequence of future light cones (constant $\mathbb{T}$ surfaces) is shown for the same observer. Light paths  are shown for various  interferometer configurations.  As a path crosses each light cone,   a projection of  transverse displacement is added, that accumulates  as a random walk and eventually  appears as a detectable difference in phase.  Note that although the outbound and inbound paths are spatially symmetric in laboratory time, their contribution to exotic correlation is not symmetric in measured signal lag, because  the opposite sign for the radial part of their propagation changes the rate at which they cross outbound light cones and thereby accumulate variance. On the right, the  red path is shown at two different beamsplitter reflection times, along with the corresponding range of integration for time lag $\tau_0$.}
    \label{fig:layers}
\end{figure}

\subsection{Effect of Exotic Displacements on the Signal}

\label{sec:effect_on_observable}

We model the effect of exotic displacements on an idealized interferometer.
Propagating light encounters a beamsplitter and divides into two separate paths, 1 and 2, of equal expected length.  The  power at the antisymmetric port is  exquisitely sensitive to the difference in path lengths\cite{2014RvMP...86..121A}.
A time series of measurements is represented by
\begin{eqnarray}
S(t) & \equiv & S_{2}(t)-S_{1}(t).\label{eq:S}
\end{eqnarray}
Here $S_1$ and $S_2$ represent the optical path lengths (OPL) of the two arms
and $S$ represents the optical path difference (OPD), in the inertial frame of the beamsplitter, and $t$ denotes laboratory time.

We now show how the exotic fluctuations
manifest as random deviations from the classical OPL. 
The two classical
optical paths will be parameterized by propagation time
$\tau\in[0,\,\mathcal{T}]$, where  $\mathcal{T}$ is the full duration of the optical phase measurement,  the length of each path.  At
propagation time $\tau$, the classical position of a tracer photon
traversing path $i$ will be denoted by $\mathbf{x}_{i}(\tau)$. These
tracer photons do not represent actual quanta of localized energy,
but rather null propagation in the inertial frame.  At every point along optical path $\mathbf{x}_{1,2}(\tau)$,
there exists a unit vector tangent to the path, $\dot{\mathbf{x}}_{1,2}(\tau)/c\equiv d\mathbf{x}_{1,2}(\tau)/cd\tau$.
This unit vector represents the instantaneous direction of light travel
through interferometer.

The light origin, $\mathbf{x}_{1}(0)=\mathbf{x}_{2}(0)$, provides
a common reference for the optical phase measurement made by either
path. It can thus be regarded, under a relational theory, as a fixed
reference point against which all other points in  space
appear to fluctuate. For each path, the optical distance propagated
by the tracer photon over a classical  light-crossing time
is given by the path integral 
\begin{eqnarray}
S_{1,2}\left(t+\mathcal{T}\right) & = & \int_{t}^{t+\mathcal{T}}\left[\dot{\mathbf{x}}_{1,2}\left(t'-t\right)+\dot{X}_{\perp}\big(\mathbb{T}(t',\,\mathbf{x}_{1,2}(t'-t))\big)\:\boldsymbol{\hat{\theta}}\right]\cdot\frac{\dot{\mathbf{x}}_{1,2}(t'-t)}{c}\,dt'\;,\label{eq:S_single_path}
\end{eqnarray}
where 
\begin{equation}
\dot{X}_{\perp}(\mathbb{T}(t,\,\mathbf{x}))\equiv\frac{d}{dt}X_{\perp}(\mathbb{T}(t,\,\mathbf{x}))=\frac{dX_{\perp}(\mathbb{T})}{d\mathbb{T}}\:\frac{d\mathbb{T}(t,\,\mathbf{x})}{dt}
\end{equation}
is the rate of transverse exotic displacement per propagation time. These displacements occur in the
direction transverse to  radial separation from the observer's world line,
a unit vector in the interferometer plane we denote by $\boldsymbol{\hat{\theta}}$. The vector
sum of this rate and the classical velocity corresponds to the ``effective''
velocity of the tracer photon relative to the classical spatial coordinates.
 The instantaneous effect on the OPL is obtained by
taking the component of this effective velocity in the direction of
light travel.

Under the variable substitution $\tau\equiv t'-t$, eq. \ref{eq:S_single_path}
becomes 
\begin{eqnarray}
S_{1,2}\left(t+\mathcal{T}\right) & = & \int_{0}^{\mathcal{T}}\left[\dot{\mathbf{x}}_{1,2}\left(\tau\right)+\dot{X}_{\perp}\big(\mathbb{T}(t+\tau,\,\mathbf{x}_{1,2}(\tau))\big)\:\boldsymbol{\hat{\theta}}\right]\cdot\frac{\dot{\mathbf{x}}_{1,2}(\tau)}{c}\,d\tau\\
 & = & \int_{0}^{\mathcal{T}}\left[\dot{\mathbf{x}}_{1,2}\left(\tau\right)+\frac{dX_{\perp}(\mathbb{T})}{d\mathbb{T}}\,\frac{d\mathbb{T}}{d\tau}(\tau,\,\mathbf{x}_{1,2}(\tau))\:\boldsymbol{\hat{\theta}}\right]\cdot\frac{\dot{\mathbf{x}}_{1,2}(\tau)}{c}\,d\tau\\
 & = & c\mathcal{T}+\int_{0}^{\mathcal{T}}\frac{dX_{\perp}(\mathbb{T})}{d\mathbb{T}}\:\mathcal{P}_{1,2}(\tau)\,d\tau
\end{eqnarray}
Here we use the fact that $d\mathbb{T}/dt'=d\text{\ensuremath{\mathbb{T}}}/d\tau$
and independent of $t$, referring to the definition in eq. \ref{eq:BSlightconetime},
and define
\begin{align}
\mathcal{P}_{1,2}(\tau) & \equiv\left[\frac{d\mathbb{T}}{d\tau}(\tau,\,\mathbf{x}_{1,2}(\tau))\right]\left[\boldsymbol{\hat{\theta}}\cdot\frac{\dot{\mathbf{x}}_{1,2}(\tau)}{c}\right]\\
 & =\left(1-\frac{1}{c}\frac{d|\mathbf{x}_{1,2}|}{d\tau}\right)\left(\boldsymbol{\hat{\theta}}\cdot\frac{1}{c}\frac{d{\mathbf{x}}_{1,2}}{d\tau}\right)\\
 & =\left(1-\frac{v_{r}}{c}\right)_{1,2}\left(\frac{v_{\theta}}{c}\right)_{1,2}
\end{align}
as a projection factor that captures all the dependence on the light
path of a specific interferometer configuration. The meaning of this
factor becomes conceptually clear when we write $\mathbf{x}=(r,\,\theta,\,z)$
in polar coordinates, and denote $\dot{\mathbf{x}}=(\dot{r},\,r\dot{\theta},\,\dot{z})=(v_{r},\,v_{\theta},\,v_{z})$.
The first half, $d\text{\ensuremath{\mathbb{T}}}/d\tau$, is the rate
of spacetime propagation relative to outgoing radial null surfaces
($r/ct=1)$, or equivalently, the number of independently fluctuating
light cones that the light path slices through per propagation time
(see Figure \ref{fig:layers}). Its value is 0 when the propagation
is purely radial in the outgoing direction, 1 when it is purely angular,
and 2 when it is purely radial in the incoming direction. This part
is a projection in the 1+1D space of $(t,\,r)$. On the other hand,
the second half, $\hat{\boldsymbol{\theta}}\cdot\dot{\mathbf{x}}/c$,
is a projection in the 2D space of $(r,\,\theta)$. It is simply the
component of the exotic displacement that is spatially tangential
to the light path.

In this form, the spatial fluctuations accumulated over a measurement
duration can be clearly seen to manifest as a deviation from the classical
OPL. The difference of optical distances along two paths  then yields
the exotic effect on the OPD, 
\begin{eqnarray}
S\left(t+\mathcal{T}\right) & = & \int_{0}^{\mathcal{T}}\frac{dX_{\perp}}{d\mathbb{T}}\big(\mathbb{T}(t+\tau,\,\mathbf{x}_{2}(\tau))\big)\;\mathcal{P}_{2}(\tau)\,d\tau-\int_{0}^{\mathcal{T}}\frac{dX_{\perp}}{d\mathbb{T}}\big(\mathbb{T}(t+\tau,\,\mathbf{x}_{1}(\tau))\big)\;\mathcal{P}_{1}(\tau)\,d\tau\label{eq:OPD_final}
\end{eqnarray}
The geometrical projection
factor $\mathcal{P}_{1,2}(\tau)$ determines the statistical
response of an instrument of arbitrary geometry.

\subsection{Single-Interferometer Statistics}

\label{sec:single_ifo_stats}

Eq. \ref{eq:OPD_final} represents OPD measurements made over an interval
of time as a set of random variables indexed by measurement time $t$.
It is a straightforward, if somewhat complicated exercise to calculate the general statistical
moments of this measurement set in the presence of exotic spatial
fluctuations.


The expected value of an OPD measurement is 
\begin{eqnarray}
\big\langle S(t)\big\rangle & = & \Bigg\langle\sum_{a}^{1,\,2}(-1)^{a}\int_{0}^{\mathcal{T}}\frac{dX_{\perp}}{d\mathbb{T}}\big(\mathbb{T}(\tau+(t-\mathcal{T}),\,\mathbf{x}_{a}(\tau))\big)\;\mathcal{P}_{a}(\tau)\,d\tau\Bigg\rangle\\
 & = & \sum_{a}^{1,\,2}(-1)^{a}\int_{0}^{\mathcal{T}}\Big\langle\frac{dX_{\perp}(\mathbb{T})}{d\mathbb{T}}\Big\rangle\,\mathcal{P}_{a}(\tau)\,d\tau\\
 & = & 0\;,
\end{eqnarray}
using the additive separability of expectation values. The zero mean
of the measurement is seen to be a direct consequence of the zero-mean
process generating the spatial fluctuations, independent of the geometry
of the apparatus.



The autocovariance of two OPD measurements separated in time by $\tau_{0}$
is 
\begin{eqnarray}\label{autocov}
C_{SS}(\tau_{0}\,|\,\ell_{P}) & \equiv & \big\langle S(t)\,S(t+\tau_{0})\big\rangle-\big\langle S(t)\big\rangle\,\big\langle S(t+\tau_{0})\big\rangle\\
 & = & \sum_{a,\,b}^{1,\,2}(-1)^{a+b}\Bigg\langle\int_{0}^{\mathcal{T}}\frac{dX_{\perp}}{d\mathbb{T}}\big(\mathbb{T}(\tau'+(t-\mathcal{T}),\,\mathbf{x}_{a}(\tau'))\big)\;\mathcal{P}_{a}(\tau')\,d\tau'\nonumber \\ \label{autocov}
 &  & \qquad\qquad\;\;\times\;\int_{\tau_{0}}^{\mathcal{T}+\tau_{0}}\frac{dX_{\perp}}{d\mathbb{T}}\big(\mathbb{T}(\tau''+(t-\mathcal{T}),\,\mathbf{x}_{b}(\tau''-\tau_{0}))\big)\;\mathcal{P}_{b}(\tau''-\tau_{0})\,d\tau''\Bigg\rangle\\
 & = & \sum_{a,\,b}^{1,\,2}(-1)^{a+b}\int_{0}^{\mathcal{T}}d\tau'\,\mathcal{P}_{a}(\tau')\int_{\tau_{0}}^{\mathcal{T}+\tau_{0}}d\tau''\,\mathcal{P}_{b}(\tau''-\tau_{0})\nonumber \\
 &  & \qquad\qquad\;\;\times\;\Big\langle\frac{dX_{\perp}}{d\mathbb{T}}\big(\mathbb{T}(\tau',\,\mathbf{x}_{a}(\tau'))\big)\,\frac{dX_{\perp}}{d\mathbb{T}}\big(\mathbb{T}(\tau'',\,\mathbf{x}_{b}(\tau''-\tau_{0}))\big)\Big\rangle\\
 & = & \sum_{a,\,b}^{1,\,2}(-1)^{a+b}\int_{0}^{\mathcal{T}}d\tau'\,\frac{d\mathbb{T}}{d\tau'}(\tau',\,\mathbf{x}_{a}(\tau'))\,\Theta_{a}(\tau')\int_{\tau_{0}}^{\mathcal{T}+\tau_{0}}d\tau''\,\frac{d\mathbb{T}}{d\tau''}(\tau'',\,\mathbf{x}_{b}(\tau''-\tau_{0}))\,\Theta_{b}(\tau''-\tau_{0})\nonumber \\
 &  & \qquad\qquad\;\;\times\;\Big\langle\frac{dX_{\perp}}{d\mathbb{T}}\big(\mathbb{T}(\tau',\,\mathbf{x}_{a}(\tau'))\big)\,\frac{dX_{\perp}}{d\mathbb{T}}\big(\mathbb{T}(\tau'',\,\mathbf{x}_{b}(\tau''-\tau_{0}))\big)\Big\rangle \label{autocov2}
 \end{eqnarray}
The first step (Eq. \ref{autocov}) follows from the vanishing mean just derived; in the second step, the statistical terms are collected together; and in the third step (Eq. \ref{autocov2}), projection factors are separated to reduce the integration to a single light cone time variable. 
We then take a final step (Eq. \ref{autofinal}) that invokes  the covariance structure of the displacements. The integral is evaluated only in the Planck-size bins where the covariance is not zero, a criterion that determines  an inverse function that maps light cone time onto the laboratory time of the corresponding part of the measured signal:
\begin{eqnarray} 
 C_{SS}(\tau_{0}\,|\,\ell_{P})  \approx & \sum\limits_{a,\,b}^{1,\,2}(-1)^{a+b}\;c\,\ell_{P}\int_{|\tau_{0}|}^{\mathcal{T}}\Theta_{a}(\xi_{a}(\mathbb{T},\,0))\,\Theta_{b}(\xi_{b}(\mathbb{T},\,|\tau_{0}|)-|\tau_{0}|)\,d\mathbb{T},\label{autofinal}
\end{eqnarray}
where we separate out just the spatial part of the projection factor
$\mathcal{P}_{1,2}(\tau)$ as
\begin{equation}
\Theta_{1,2}(\tau)\equiv\boldsymbol{\hat{\theta}}\cdot\frac{\dot{\mathbf{x}}_{1,2}(\tau)}{c}
\end{equation}
and $\xi_{1,2}$ is an inverse function for $\mathbb{T}(t,\,\mathbf{x})$
defined such that
\begin{align}
\tau'=\xi_{1,2}(\mathbb{T},\,0)\qquad\textrm{when}\qquad & \mathbb{T}=\mathbb{T}(\tau',\,\mathbf{x}_{1,2}(\tau'))\\
\tau''=\xi_{1,2}(\mathbb{T},\,\tau_{0})\quad\;\;\,\textrm{when}\qquad & \mathbb{T}=\mathbb{T}(\tau'',\,\mathbf{x}_{1,2}(\tau''-\tau_{0})).\label{eq:inverse}
\end{align}

The inverse function is well-defined everywhere except when $\mathbf{x}_{1,2}$
is purely radial in the outgoing direction, where $\mathbb{T}$ is
constant over a null path (everywhere else, $\mathbb{T}$ is monotonically
increasing). However, over those segments, $\Theta_{1,2}=0$, making
those parts irrelevant.
The last step (Eq. \ref{autofinal}) should be a highly accurate approximation mathematically; the Planck scale is almost infinitesimal for practical purposes.
However, since we treat the covariance structure $\left\langle dX_{\perp}/d\mathbb{T}\;\,dX_{\perp}/d\mathbb{T}\right\rangle $
as providing bins of Planck length width, our 
spacetime is not continuous but discrete at
the Planck scale.  Formally the integral picks out a single value of $\Theta_{1,2}$
for each bin in $\mathbb{T}$, instead of treating $\Theta_{1,2}$
as slowly-varying continuous functions. For nearly or exactly radial propagation, the difference can be important in numerical computations, but these regions are suppressed by near-zero values of $\Theta_{1,2}$.
The mapping between $\mathbb{T}$ and $\tau$ is nonlinear and dependent
on $\mathbf{x}_{1,2}$, which means that in practice, for a specific
interferometer configuration, the range of intervals for $\tau'$
and $\tau''$  map nontrivially to an integration over $\mathbb{T}$.
We show examples of this below. 

 Equations (\ref{autocov}) to  (\ref{autofinal}) specify the effect of Planck scale correlations on the signal of any interferometer. In general, the covariance does not vanish but behaves approximately like a Planck random walk for some intervals of time, for a layout where the light paths have a projected transverse component. The autocovariance identically vanishes for separations $|\tau_{0}|$
larger than the duration over which an individual measurement has
nonzero geometrical projection $\mathcal{P}_{1,2}(\tau)$ (see Figure \ref{fig:layers}).



Under the Wiener-Khinchin theorem, an equivalent, frequency-space
representation of the autocovariance is the power spectral density
(PSD). The PSD is defined as the Fourier transform of the autocovariance,
\begin{eqnarray}\label{autospec}
\widetilde{C_{SS}}(f\,|\,\ell_{P}) & \equiv & 2\int_{-\infty}^{\infty}C_{SS}(\tau_{0}\,|\,\ell_{P})\,e^{-i2\pi f\tau_{0}}\,d\tau_{0}\\
 & = & 4\int_{0}^{\infty}C_{SS}(\tau_{0}\,|\,\ell_{P})\,\cos\left(2\pi f\tau_{0}\right)\,d\tau_{0}\;,
\end{eqnarray}
where the second equality follows from the fact that $C_{SS}(\tau_{0}\,|\,\ell_{P})=C_{SS}(-\tau_{0}\,|\,\ell_{P})$.
This PSD, written in the so-called engineering convention, is defined
only for positive frequencies, in which the power contained in the
redundant negative frequencies is folded via the multiplicative prefactor
of two.

\section{Projections for Experimental Layouts}

\label{sec:phase_II_design}

\begin{figure}
\centering \includegraphics[height=0.96\textheight]{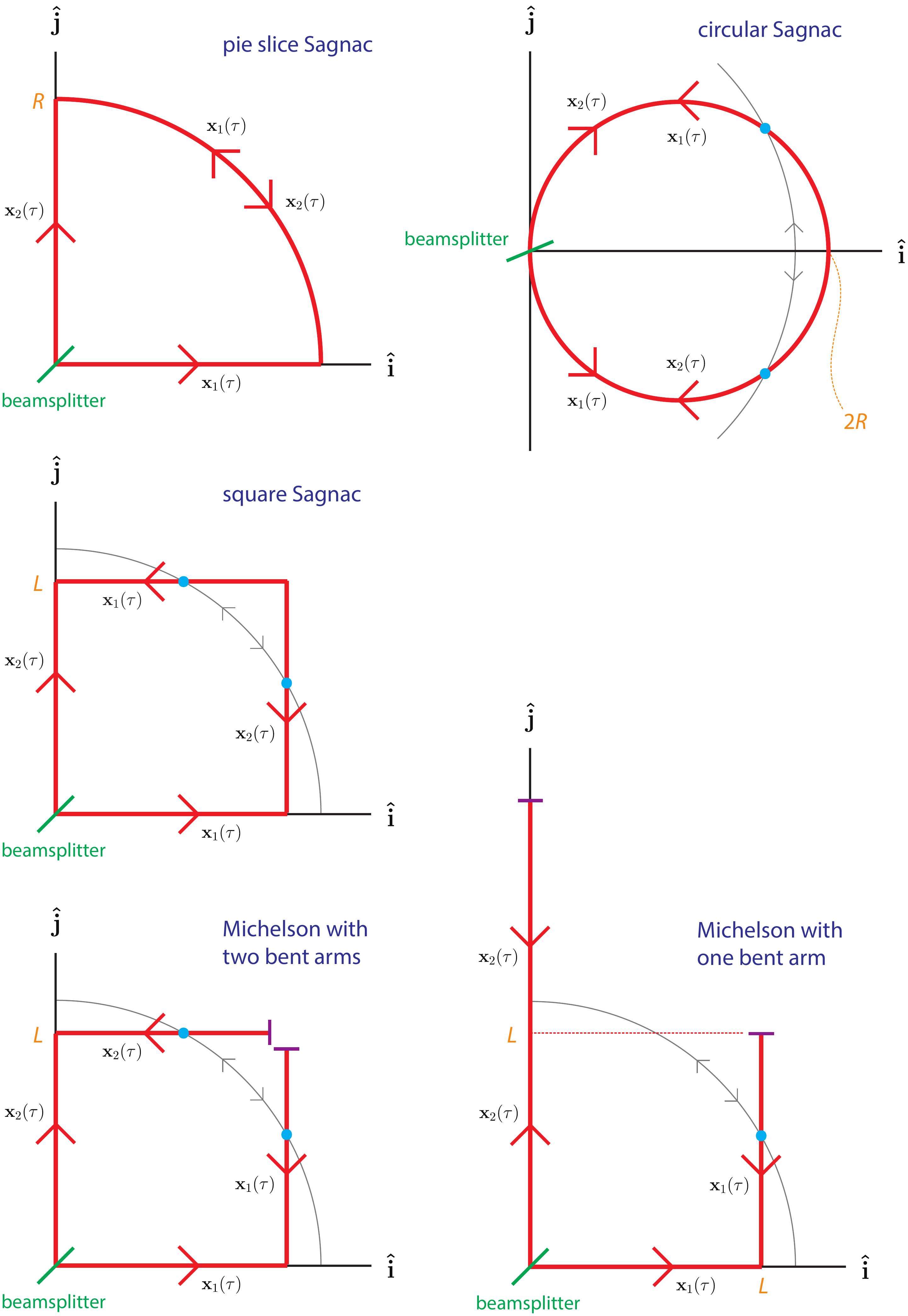}\caption{Schematics of the optical paths in various interferometer setups where
the light phase is sensitive to exotic rotation.}
\label{fig:interferometers} 
\end{figure}

To aid in conceptual design of experiments, and to gain insight into the character of the exotic correlations,  it is useful to survey the signal response of specific interferometer configurations (see Figure \ref{fig:interferometers}). Two features stand out as required for detectable effect: (1) a layout where light travels a substantial distance in a direction that is not purely radial with respect to the ``observer'' (that is, the beamsplitter whose position  defines the measured path difference); and (2) the ability to measure correlations on a timescale short compared with a light travel time in the interferometer arms, or equivalently over a frequency bandwidth comparable to the free spectral range.

The calculational framework above
is designed for  two  optical paths  labeled by 1 and 2 that eventually recombine.
In the Michelson configurations, a beamsplitter sends light into two separate arms. In the Sagnac configurations\cite{RevModPhys.39.475,Schreiber2014859}, the split light travels in opposite directions around the same path. 
In a Sagnac system, we consider tracer photons
propagating in opposite directions and label their optical paths by
1 and 2. The formalism applies entirely unchanged in both cases.
Idealized configurations, the ``pie slice
Sagnac'' and the ``circular Sagnac,''  have curved light paths that can be considered as limiting cases of many short segments. They
 are presented as thought
experiments to demonstrate statistical characteristics of the
covariance structure. More realistic configurations for experimental setups are a  ``square
Sagnac'' and a ``bent Michelson.'' One form of the latter will be the main
focus of the second phase of the Holometer program, as it  is technologically simpler to implement and allows a reliable null configuration.

Throughout this section, we present our spectra normalized with respect
to a system timescale $\mathcal{T}_{0}$. This is sometimes the same
as light circulation time $\mathcal{T}$, but often, because the light
phase is not sensitive to rotational shifts when its propagation is
purely radial, the effective optical path length over which there
is nonzero geometrical coupling is smaller. $\mathcal{T}_{0}$ is
the duration of time a tracer photon spends within these segments\textemdash{} for example, for the bent Michelson configuration, $\mathcal{T}_{0}$
is round trip time within the bent portion of the arm, whereas $\mathcal{T}$
is the total round trip time.

\subsection{Michelson Interferometer with Straight Arms}

A standard
Michelson interferometer, commonly used in gravitational wave observatories
and in the initial phase of the Holometer experiment, has no sensitivity
to the exotic rotational effect, due to the fact that the light propagation is
always radial with respect to the reference observer world line at
the beamsplitter, so that the geometrical coupling $\Theta\left(\tau\right)$
vanishes over the whole light path. Some spatial element of angular propagation
is necessary to have nonzero response.
This property provides a useful way to design a null experiment to calibrate environmental noise and systematic errors.

\subsection{Pie Slice Sagnac Interferometer}

This is the simplest configuration to understand, with a perfectly
linear time-domain spectrum that is conceptually straightforward to
derive. As shown in Figure \ref{fig:interferometers}, it consists
of two purely radial segments of equal length \textit{R}, both of
them with zero geometrical coupling $\Theta\left(\tau\right)$ to
rotational jitter, connected by a circular arc that spans a quarter
circle of radius \textit{R}, over which the light propagation is purely
angular and the projection factors $\mathcal{P}_{1,2}(\tau)$ take
values of $\pm1$. The completed loop therefore has a system timescale
of just $\mathcal{T}_{0}=\frac{1}{2}\pi R/c$ (despite the light circulation
time being $\mathcal{T}=(\frac{1}{2}\pi+2)R/c$), and the total accumulated
correlation as a function of time lag $\tau_{0}$ is:
\begin{equation}
C_{SS}(\tau_{0}\,|\,\ell_{P})=\begin{cases}
2^{2}\,c\,\ell_{P}\,\mathcal{T}_{0}\,(1-|\tau_{0}|/\mathcal{T}_{0})\;, & 0<|\tau_{0}|<\mathcal{T}_{0}\\
0\;, & \textrm{otherwise}
\end{cases}
\end{equation}
where the factor of $2$ comes from the two optical paths each picking
up the same jitter. The time-domain signal is an autocovariance of
the  OPD fluctuations, so the factor is squared (see Eq. \ref{autocov}). As the time lag $\tau_{0}$ increases in
magnitude, the duration of beamsplitter light cone time $\mathbb{T}$
over which there are correlated fluctuations decreases linearly, up
to the total timespan $\mathcal{T}_{0}$ over which the light propagation
is angular and the light phase is sensitive to rotational shifts.

The frequency-domain PSD can easily be calculated analytically:
\begin{equation}
\widetilde{C_{SS}}(f\,|\,\ell_{P})=8\,c\,\ell_{P}\,\mathcal{T}_{0}^{2}\,\textrm{sinc}^{2}(\pi\,f\,\mathcal{T}_{0})
\end{equation}

The resulting spectrum is shown in Figure \ref{fig:autospectra}.  This simple model also gives a fairly close approximation to the exotic noise spectrum in the Sagnac mode\cite{Hogan:2001jn} of a triangular interferometer, such as  LISA.

\subsection{Circular Sagnac Interferometer}

Consider now an idealized Sagnac, a circular loop of radius \textit{R}
with two light paths going around it in opposite directions, this  time starting
from a beamsplitter on the circumference. The geometrical coupling
$\Theta\left(\tau\right)$ over the light path is a perfect sinusoidal
function with one single frequency, at a system timescale equal to
light circulation time $\mathcal{T}_{0}=\mathcal{T}=2\pi R/c$.
The two optical paths can be parameterized as follows (in polar coordinates,
$\mathbf{x}=(r,\,\theta)$):
\begin{equation}
\mathbf{x}_{a}(\tau)=\left(\frac{c\mathcal{T}_{0}}{\pi}\sin\frac{\pi\tau}{\mathcal{T}_{0}}\,,\;\pi(-1)^{a}\bigg(\frac{1}{2}-\frac{\tau}{\mathcal{T}_{0}}\bigg)\right)
\end{equation}
where $a=1,\,2$. We then calculate:
\begin{equation}
\Theta_{a}\left(\tau\right)=\boldsymbol{\hat{\theta}}\cdot\frac{\dot{\mathbf{x}}_{a}(\tau)}{c}=(-1)^{a-1}\sin\frac{\pi\tau}{\mathcal{T}_{0}}
\end{equation}
\begin{equation}
\mathbb{T}(\tau',\,\mathbf{x}_{1,2}(\tau))=\tau'-\frac{\mathcal{T}_{0}}{\pi}\sin\frac{\pi\tau}{\mathcal{T}_{0}}\label{eq:TvsTauCircular}
\end{equation}
and similar for $\tau''$, with $\tau=\tau''-\tau_{0}$ used as the
input. The inverse functions $\xi_{1,2}(\mathbb{T},\,\tau_{0})$ are
well-defined over the entire domain, but not in analytical form. 
The integrals for $C_{SS}(\tau_{0}\,|\,\ell_{P})$ and $\widetilde{C_{SS}}(f\,|\,\ell_{P})$
can also only be evaluated numerically. The results are shown in Figure
\ref{fig:autospectra}. 

Notably, despite the perfectly symmetrical and sinusoidal form of
the geometrical coupling, the nonlinear mapping between $\mathbb{T}$
and $\tau$ in eq. \ref{eq:TvsTauCircular} results in an asymmetric
weighting over the light circulation time, because there is more accumulated
fluctuation when the light path slices through more layers of future
light cones per time.  This asymmetric response might seem counterintuitive, but it is an important feature of the theory that is a natural consequence of the arrow of time associated with quantum measurement (see section \ref{sec:timeasymmetry}). If one naively assumes a covariance
structure that is linearly dependent on $\tau$ instead of defining
$\mathbb{T}$ on future light cones, the resulting PSD goes negative
at certain frequencies, which means that it cannot represent a real-valued
physical observable undergoing a wide-sense stationary random process\cite{oppenheim2016signals}.

\subsection{Square Sagnac Interferometer}


The calculations for the device response and spectra expected in a
square Sagnac closely follow those of a configuration to be discussed
in the next section, a Michelson interferometer with one or two arms
bent $90\textdegree$ inward at the midpoint (see Figure \ref{fig:interferometers}).
Since we are positing a Planckian jitter that is stochastic and covariant
on light cones, the magnitude of the total accumulated correlated fluctuation
per optical path (1 or 2) is exactly the same for Sagnac and Michelson layouts at zero time lag $\tau_{0}$,
despite the difference of a round trip versus a complete loop.  The
mapping between $\mathbb{T}$ and $\tau$ is also identical (see Figure
\ref{fig:layers}). The geometrical coupling $\Theta\left(\tau\right)$
differs by a sign change halfway along the light path, which does
significantly change the behavior of the cross correlation at nonzero
time lag and thereby the distribution of fluctuation power in the
frequency domain. The resulting $C_{SS}(\tau_{0}\,|\,\ell_{P})$ and $\widetilde{C_{SS}}(f\,|\,\ell_{P})$ are shown 
in Figure \ref{fig:autospectra}.

\subsection{Bent Michelson Interferometer}

It is easy to see the schematic
similarities between a square Sagnac interferometer and a Michelson
with both arms folded inward to form a square. The key difference
is whether the light makes complete loops around the square or reflects
back halfway through the light path. For the classical Sagnac effect, as in 
classical rotations measured with optical gyroscopes, this difference
matters, because the time-symmetric round trip light path within a
bent Michelson would show zero coupling to that effect. But 
the exotic rotational fluctuations are stochastic in time and follow  the same covariance structures defined on
concentric future light cones around an observer at the beamsplitter, so
they accumulate total variance at  exactly the same rate
in both devices, although that variance is distributed differently in measured time and frequency.

Sagnac interferometers are routinely operated  as optical gyroscopes,
 but not with the  sensitivity required at high frequency to measure  exotic rotational  correlations.  
For technical reasons, the Holometer program plans instead to  to search for exotic rotational correlations with  a pair of  Michelson interferometers with only one bent arm. 
We describe this set up in more explicit detail than
the other examples.  
We also evaluate the displayed solution for definite
physical dimensions that approximate the real apparatus. One \textquotedblleft north\textquotedblright{}
arm is $2L=39$ meters long, and the other \textquotedblleft east\textquotedblright{}
arm is bent in the middle with a folding mirror at $L=19.5$ meters from
the beamsplitter. The reference design is shown in Fig. \ref{fig:interferometers}. 
Since the straight arm has zero coupling to the effect, this
device has $1/2^{2}$ the amount of accumulated variance observable
in a Michelson with two bent arms.

Define unit vectors $\mathbf{\hat{i}}$, $\mathbf{\hat{j}}$ in the
east and north directions respectively. Adopting a beamsplitter-centered
coordinate system, the classical round-trip light paths through the
east and north arms can be parameterized as 
\begin{eqnarray}
\mathbf{x}_{1}(\tau)=\begin{cases}
c\tau\mathbf{\hat{i}}\;, & 0<\tau<T\\
L\mathbf{\hat{i}}+(c\tau-L)\mathbf{\hat{j}}\;, & T<\tau<2T\\
L\mathbf{\hat{i}}+(3L-c\tau)\mathbf{\hat{j}}\;, & 2T<\tau<3T\\
(4L-c\tau)\mathbf{\hat{i}}\;, & 3T<\tau<4T
\end{cases} & \qquad\qquad & \mathbf{x}_{2}(\tau)=\begin{cases}
c\tau\mathbf{\hat{j}}\;, & 0<\tau<2T\\
(4L-c\tau)\mathbf{\hat{j}}\;, & 2T<\tau<4T
\end{cases}
\end{eqnarray}
where $T=L/c$ and the arm-segment length $L=19.5$m. It is straightforward
to calculate the geometrical couplings: 
\begin{align}
\Theta_{1}\left(\tau\right) & =\boldsymbol{\hat{\theta}}\cdot\frac{\dot{\mathbf{x}}_{1}(\tau)}{c}=\begin{cases}
0\;, & 0<\tau<T\\
\Theta_{\shortrightarrow}(\tau)\equiv\frac{1}{\sqrt{1+(\tau/T-1)^{2}}}\;, & T<\tau<2T\\
\Theta_{\shortleftarrow}(\tau)\equiv-\frac{1}{\sqrt{1+(3-\tau/T)^{2}}}\;, & 2T<\tau<3T\\
0\;, & 3T<\tau<4T
\end{cases}\label{eq:east-arm-theta}\\
\Theta_{2}\left(\tau\right) & =\boldsymbol{\hat{\theta}}\cdot\frac{\dot{\mathbf{x}}_{2}(\tau)}{c}=0\label{eq:north-arm-theta}
\end{align}
where we see that the system timescale over which the light phase
is sensitive to rotational shifts is actually $\mathcal{T}_{0}=2L/c$,
instead of the light circulation time $\mathcal{T}=4L/c$. Here the arrows
label the outgoing and incoming halves of the light round trip within
the bent portion of the arm (the negative sign on the return half
of eq. \ref{eq:east-arm-theta} is the only difference we have with
the square Sagnac calculations, because there the light completes
the loop in one direction).

\begin{figure}
\begin{centering}
\centering \includegraphics[width=0.98\linewidth]{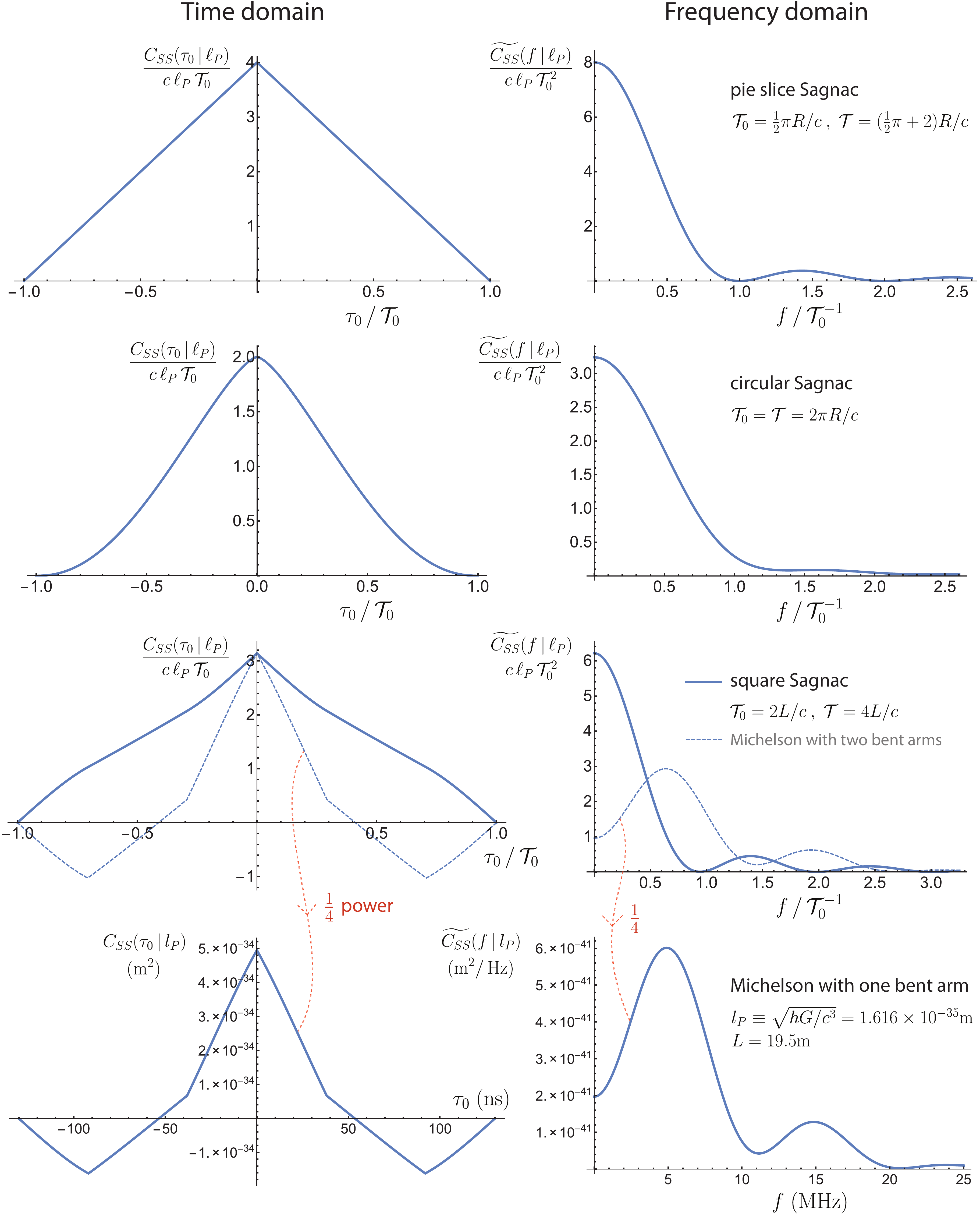}
\par\end{centering}
\caption{The time- and frequency-domain autocovariance of signals in various
interferometer setups, normalized to their system scales. For the
bent Michelson, predictions in physical units are presented, with nominal standard
Planck length normalization and the approximate dimensions of an experiment
being commissioned. The precise coherence scale is considered a parameter
to be fixed by experiment. Note locations of time domain inflections from mirror reflections, as discussed
in Appendix \ref{sec:bentspectra}. In the square Sagnac, they occur in the second derivative.}
\label{fig:autospectra} 
\end{figure}

We also derive the mapping between propagation time and beamsplitter
light cone time in the bent ``east'' arm, shown in Figure \ref{fig:layers}:
\begin{equation}
\mathbb{T}(\tau'',\,\mathbf{x}_1(\tau''-\tau_{0}))=\begin{cases}
\tau_{0} & 0<\tau''-\tau_{0}<T\\
\mathbb{T}_{\shortrightarrow}(\tau'',\,\tau_{0})\equiv\tau''-\sqrt{T^{2}+(\tau''-\tau_{0}-T)^{2}} & T<\tau''-\tau_{0}<2T\\
\mathbb{T}_{\shortleftarrow}(\tau'',\,\tau_{0})\equiv\tau''-\sqrt{T^{2}+(\tau''-\tau_{0}-3T)^{2}} & 2T<\tau''-\tau_{0}<3T\\
\tau_{0}+2(\tau''-\tau_{0}-2T) & 3T<\tau''-\tau_{0}<4T
\end{cases}\label{eq:Ttau}
\end{equation}
with a similar expression for $\tau'$, with $\tau_{0}=0$. 
These can then be used to evaluate the autocovariance $C_{SS}(\tau_{0}\,|\,\ell_{P})$
and autospectrum $\widetilde{C_{SS}}(f\,|\,\ell_{P})$, from Eqs. (\ref{autocov2}) and (\ref{autospec}). The numerical
solution is shown in Fig. \ref{fig:autospectra} for $\ell_{P}=l_{P}$.
A more detailed discussion of the origin of the segmented structure
is given in Appendix \ref{sec:bentspectra}. 

Interestingly, this spectrum has a nonzero DC response. This might
seem counterintuitive, since the light path follows perfectly time-symmetric
round trips (unlike in the Sagnac examples): a bent Michelson  has no signal response to classical constant rotation. But due to the nonlinear
mapping in eq. \ref{eq:Ttau}, the outward trip picks up much less
jitter than the portion returning inward, when the light path slices
through more  layers of  future
light cones per unit time.  The fluctuations 
 average to zero, but the device response to exotic twists appears asymmetric. As discussed below (section \ref{sec:timeasymmetry}), this
is a consequence of asymmetry between  future and past in the preparation  of states and decoherence associated with a measurement.
Since the frequency spectrum approaches a constant
value at zero frequency,  the mean square displacement decreases approximately linearly for averaging times
 longer than the apparatus size, so it  goes to zero in the limit of a long time average.
The dominant contribution to the total displacement still comes from frequencies
on the order of the free spectral range.  

With Planck length normalization, the predicted signal is detectable.
Assuming sensitivity similar to that already achieved with the Fermilab
Holometer\cite{Holo:PRL,Holo:Instrument}, a highly significant detection is expected after several
hours of integration. With several hundred hours of data, the program
should be able to probe an order of magnitude lower.

\section{Spatial Entanglement and  Cross-Correlation}

\label{sec:spatial_entanglement}

Consider a collection of tangent  2-spheres with a variety of different radii. They represent 2-sections of light cones corresponding to a variety of different observers' world lines.
According to the future-light-cone covariance hypothesis, these spheres all share the same exotic phase displacement in the tangent directions, and must  ``choose'' the same value of $\delta X_\perp$ to collapse the Planck subsystem at their intersection point--- the same projection of exotic twist onto any axis. Thus, covariance on causal boundaries  appears as  a sort of spooky conspiracy among many observers belonging to these light cones. How do the Planck  subsystems ``know about'' their large scale relationships with distant observers?  

The answer of course is that the relationship among  quantum subsystems is always determined by the preparation and measurement of the  whole state.  The covariance represents entanglement, which is  imposed by the assembly of subsystems into a whole system. The  projection of the state is fixed by the world line of an observer. Different observers  see different projections, but the whole system collapses in a self consistent way.  The system is not exactly classical, but has  ``spooky'' correlations. They appear to be nonlocal, with spacelike separations, but actually rigorously respect causal structure.

Indeed, the basic principle is that ``all correlations are local'':  entanglement of states leads to the same transverse  displacements for all events on a line with zero space time interval separation.
For one observer, we have expressed this statement in terms of future light cone covariance, but
the statement still applies in the case of signals cross correlated between two separate interferometers.
Exotic  fluctuations measured by nearly co-located interferometers
are expected to exhibit a high degree of correlation. The degree of entanglement
between spatial positions   can be expressed in terms of the covariance
structure on light cones.

\begin{figure}[ht]
   \centering
       \includegraphics[width=0.7\linewidth]{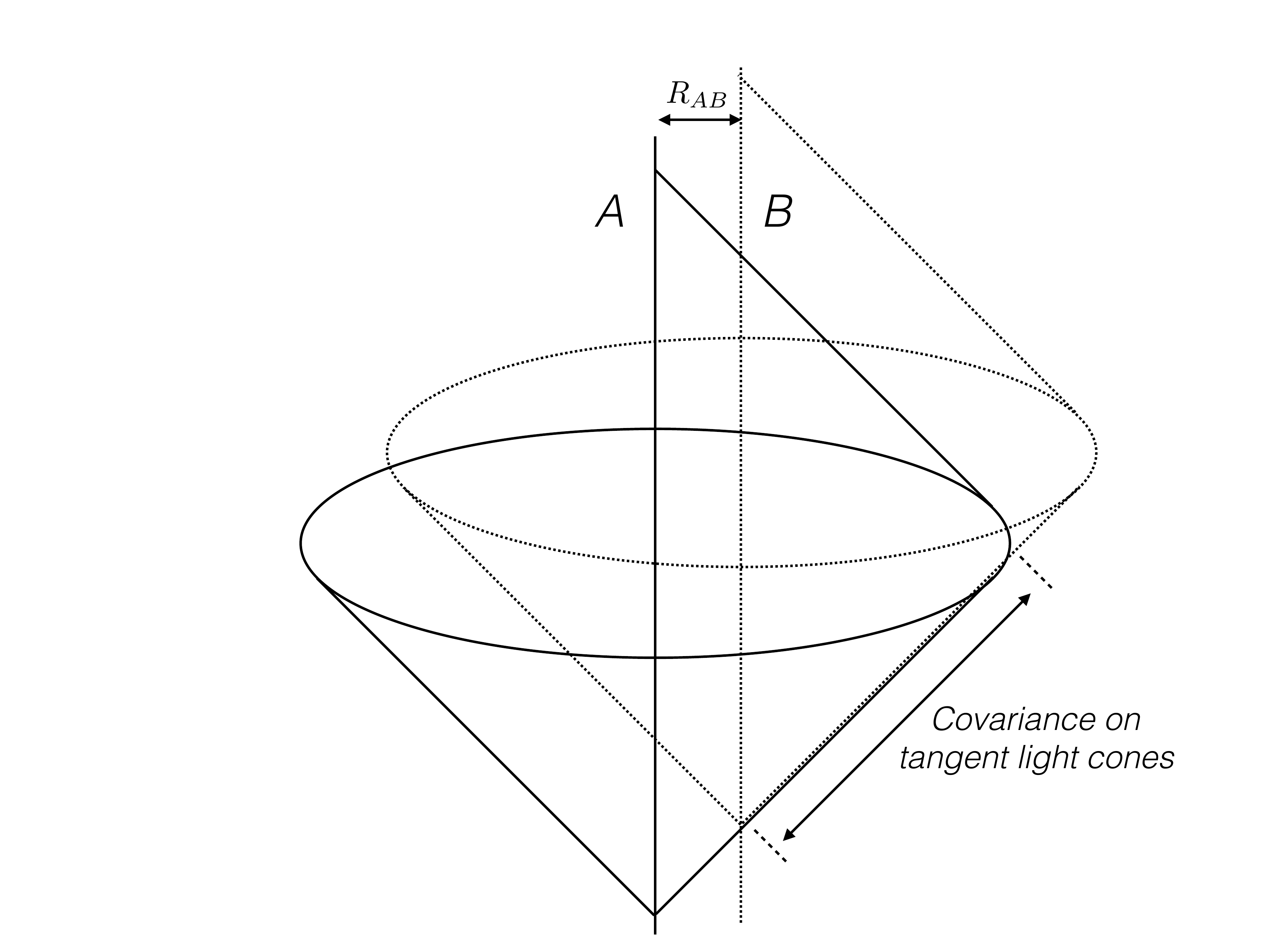}
   \caption{The entanglement of exotic rotational  displacements measured relative to two observers $A$ and $B$  is fixed by the in-common future tangent light cones. Along the spatial direction separating the observers, but only at laboratory time separations larger than their separation $R_{AB}/c$, exotic displacements for the two are the same. If  the $R_{AB}$ between beamsplitters is small compared to the region explored by their optical paths,  signals of two interferometers display significant cross correlation. Thus, exotic rotational fluctuations are highly correlated in both space and time.}
   \label{crosscovariance}
\end{figure}

Consider  the configuration
of two observer world lines $A$ and $B$,  with separation  $R_{AB}$, as  illustrated in Fig. \ref{crosscovariance}.
The displacement rates  $dX_{\perp}^{A,B}/d\mathbb{T}_{A,B}$ are  relational variables, specific to  each observer and its light cone time.
Covariance has been defined for future light cones of each observer, but the two sets of displacements are not independent of each other because they share the same emergent space-time.
The variables  refer to displacements in specific emergent directions in space: 
projections tangent to the radial vector to each observer.
For two observers $A$ and $B$, the directions  agree  along one line in space, the continuation of the segment separating the observers, where light cones are tangent.
Along this line, transverse projections are identical at the same event in the same direction:
 displacements with respect to  each of the two observers agree
 in  magnitude and direction, for events in their respective future light cones.
  
This constraint  determines the entanglement  over the rest of the space-time.
The resulting covariance can be written in terms of the two  light cone times:
\begin{equation}
\textrm{cov}\left(\frac{dX_{\perp}^{A}}{d\mathbb{T}_A}(\mathbb{T}_A),\,\frac{dX_{\perp}^{B}}{d\mathbb{T}_B}(\mathbb{T}_B)\right)=\begin{cases}
{\displaystyle \left(\frac{\ell_{P}}{t_{P}}\right)^{2},} & \left| \mathbb{T}_A - \mathbb{T}_B\right|<\frac{1}{2}t_{P}\\
0\;, & \textrm{otherwise.}
\end{cases}\label{eq:entanglement}
\end{equation}
Analogously to the single-interferometer
case (Eq. \ref{autocov}), the cross-interferometer statistics of  two measurement sets
can now be calculated by using the covariance in Eq. (\ref{eq:entanglement}):
\begin{eqnarray}
C_{AB}(\tau_{0}\,|\,\ell_{P}) & = & \big\langle S_A(t)\,S_B(t+\tau_{0})\big\rangle\\
 & = & \sum_{a,\,b}^{1,\,2}(-1)^{a+b}\Bigg\langle\int_{0}^{\mathcal{T}}\frac{dX_{\perp}^A}{d\mathbb{T}_A}\big(\mathbb{T}_A(\tau'+(t-\mathcal{T}),\,\mathbf{x}_{A,a}(\tau'))\big)\;\mathcal{P}_{A,a}(\tau')\,d\tau'\nonumber \\
 &  & \qquad\qquad\;\;\times\;\int_{\tau_{0}}^{\mathcal{T}+\tau_{0}}\frac{dX_{\perp}^B}{d\mathbb{T}_B}\big(\mathbb{T}_B(\tau''+(t-\mathcal{T}),\,\mathbf{x}_{B,b}(\tau''-\tau_{0}))\big)\;\mathcal{P}_{B,b}(\tau''-\tau_{0})\,d\tau''\Bigg\rangle,
  \label{crosscov}
\end{eqnarray}
where the average is over a time $t$  in proper time defined by a laboratory clock world line.

Although this formula is almost the same as the autocovariance (Eq. \ref{autocov}), the final result differs from  Eq.(\ref{autofinal}),  because the separation $R_{AB}$ enters implicitly, via the covariance and the range of integration.
In laboratory time $t$, the  agreement $ \mathbb{T}_A = \mathbb{T}_B$  occurs at time 
offset of $\left | t_A-t_B \right | = R_{AB}/c$.
This slightly modifies the range of integration, which reduces the signal cross covariance below that of the autocovariance. 

\begin{figure}[ht]
   \centering
       \includegraphics[width=0.4\linewidth]{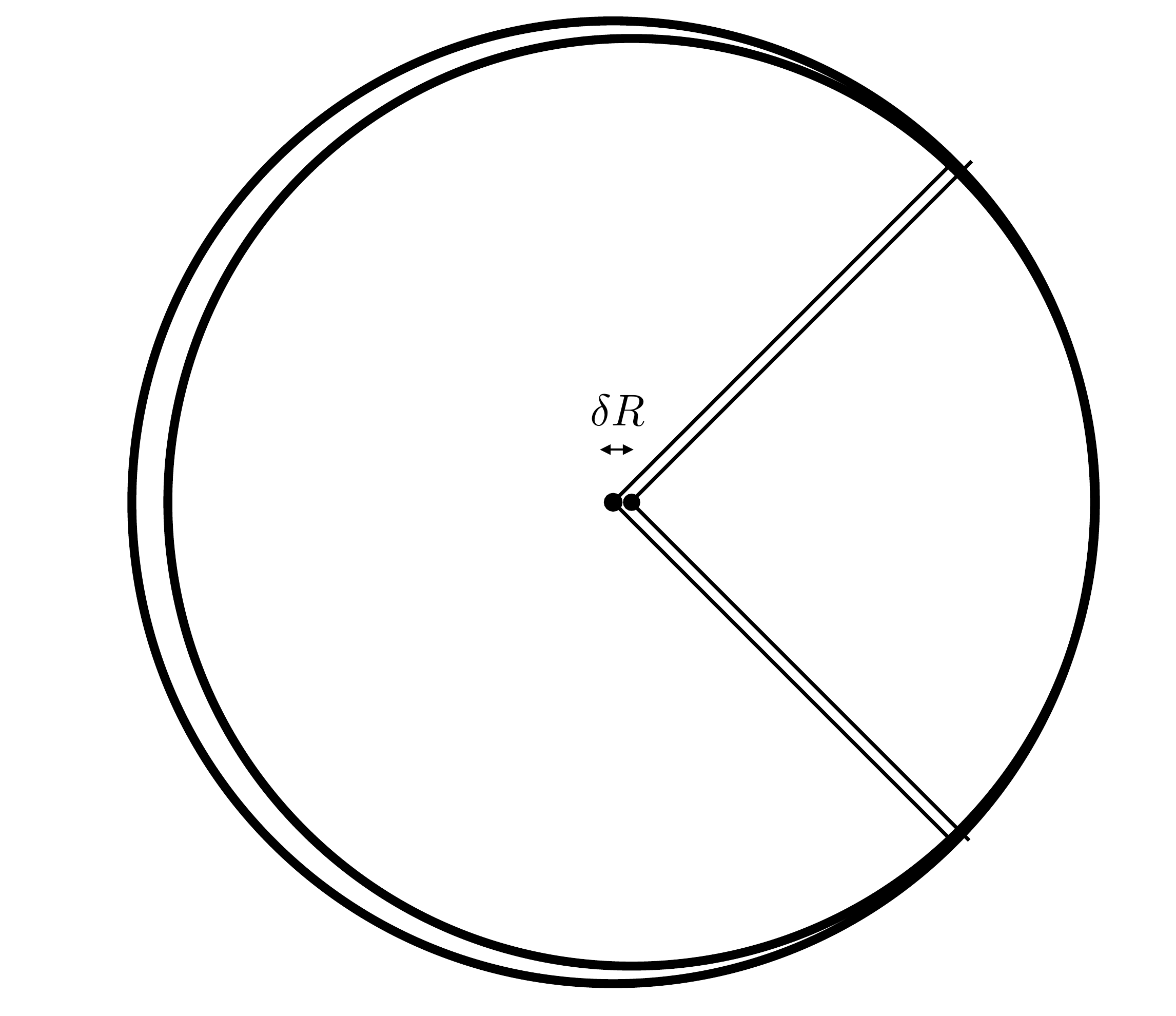}
   \caption{Slices of two tangent light cone surfaces, with slightly separated centers, at a single time in the lab frame. Because they are tangent, the exotic transverse displacements relative to their respective centers are identical everywhere on both circles.  For an offset $\delta R$ much smaller than the overall scale $R$, the   integrals  that project these displacements onto neighboring pie-slice Sagnac signals (say) differ by only a small fractional amount, of the order of $\delta R/R$. Thus, the cross correlation of two machines is close to the autocorrelation, to first order in $\delta R/R$, even for macroscopic values of $\delta R$.  Similarly, it is a good approximation to assume a constant displacement in the plane of the propagating wave fronts or reflecting optics, as long as they are small compared to the  scale of the whole system being measured.}
   \label{twoslice}
\end{figure}

However, for  two similar interferometers separated by a distance  much smaller than their size, the
non-overlapping parts of the integral due to the time offset   are small compared its total value, and the cross covariance is almost equal to the autocovariance. 
The circle or pie slice examples  illustrate this point explicitly. Consider two pie slice interferometers with a small radial displacement offset $ R_{AB} = \delta R <<R$. The offset can still be macroscopic, much larger than wavelength, beam width or even mirror size. Since the light cones  are almost concentric and the projection factors are geometrically identical except for this fractionally small offset, the nonvanishing part of the correlation integral  changes by only a small factor, $\delta C/C= (C_{AB}- C_A)/C_A \approx -\delta R/R$ (see Fig. \ref{twoslice}). 


Thus, the signals measured by neighboring
interferometers correlate in time purely due to the spatial proximity of the
instruments, even if their optical paths do not coincide.
This behavior makes intuitive sense: they  measure the exotic rotation of almost the same emergent space-time volume  at almost the same time, in  almost the same way. The exotic displacements on light cones close to the tangent cone are almost  the same, because the directions to the beamsplitters at nearly the same and place nearly agree.
This property is important in the  design of experiments, since it allows them to take advantage of cross correlation to eliminate many spurious sources of noise and signal contamination.


\section{Comments and Interpretation}

\subsection{Discrete Qubit  Model of Light Cone Twists}\label{sec:qubitmodel}

The statistical model shows how space-time position relationships can emerge statistically
if  covariance properties of the random variable $\delta X_\perp$ follow the causal structure defined by an observer's light cones. 
It is useful to sketch  an explicit example  that shows how random variables with these covariance properties can emerge from a discrete quantum system.
This simple system shows how the intrinsically quantum properties of  superposition and collapse or measurement at the Planck scale map  onto our classical model. In particular, it shows how the basic assumptions of our statistical model follow from a new physical interpretation of a standard quantum spin system--- essentially, exact causal symmetry and statistical Lorentz invariance, recognized by  relabeling of coordinates. In this example the quantum system is the same simple two-state qubit widely studied in the literature on quantum computing and entanglement,  except for the physical reinterpretation of the quantum observables.




An approximation to a  world line  emerges from a discrete sequence of events.
Each discrete element  maps onto a  light cone.
The light cone in the classical world is 3 dimensional, with 2 space dimensions and one null dimension that corresponds to both time and radial spatial separation in the observer's frame. To represent this 3D system in  a discrete quantum system, we adopt  displacement operators $\delta \hat X_i$ that obey the standard  spin algebra 
\begin{equation}\label{paulispin}
[\delta \hat X_i, \delta \hat X_j] = i l_P \epsilon_{ijk} \delta \hat X_k
\end{equation}
where $\epsilon_{ijk}$ denotes the antisymmetric tensor.
Algebraically the $\delta \hat X_i$ are exactly equivalent to standard Pauli matrices $\sigma_i$ used for quantum angular momentum and spin, but expressed in units of Planck length instead of Planck's constant, that is, $\hbar\rightarrow l_P$. The eigenvalues for the the standard basis states are then in physical units with modulus $ l_P$.

 A choice of direction to project a light cone state  is analogous to choosing an axis to project the state of a spin system. In our application, the radial direction from the observer has a special classical significance, because of  its special relationship  with emergent time.
With no loss of generality, we   choose to identify the  $i,j,k= 3$-axis with the  radial direction in the classical correspondence limit, which is also unique   the null or time direction of the light cone from a given event, in the  observer's frame. 
This projection is represented by the operator that  corresponds to the standard Pauli matrix label $i,j,k=3$ or sometimes simply $Z$.
The other two axes  (the conventional  Pauli matrix labels $i,j,k=1,2$ or simply $X,Y$) map onto  left/right displacements in transverse directions, about each independent classical rotation axis. 
With these choices, null values for the position observables correspond to the classical  inertial frame of the observer, and other values represent deviations from that.

Using standard notation, the state of the system is represented by a  vector  $|\psi \rangle$ in a Hilbert space.  
The basis states along the coordinate axes represent different spatial projections of a two state system:  a single qubit of  quantum information.
For our  2+1 D application we modify the standard notation. We  adopt new labels $|L_1\rangle,|R_1\rangle$ or  $|L_2\rangle,|R_2\rangle$  for the basis states that represent ``left'' or ``right'' twists around the $i,j,k= 1,2$ axes respectively:  these correspond to our exotic transverse displacements.  (In quantum computing the  $X,Y$ axis states are commonly labeled   $|+\rangle,|-\rangle$ and  $|+i\rangle,|-i\rangle$ respectively).
We adopt new labels $|\uparrow\rangle, |\downarrow\rangle$ to label the basis states of $i,j,k=3$, the null or radial direction. (These $Z$ axis state  are commonly labeled  $|0\rangle,|1\rangle$.) 
The algebra determines how the eigenstates in each projection can be expressed as  superpositions of states in the other directions. For example: 
\begin{equation}
| \uparrow \rangle  = { 1\over \sqrt{2}} (| L_1\rangle + | R_1\rangle) = { 1 \over \sqrt{2}} (| L_2\rangle + | R_2\rangle), \ \qquad | \downarrow \rangle  = { 1\over \sqrt{2}} (| L_1\rangle - | R_1\rangle) = { 1 \over \sqrt{2}} (| L_2\rangle - | R_2\rangle)
\end{equation}
\begin{equation}
|L_1 \rangle  = { 1\over \sqrt{2}} (| \uparrow \rangle + | \downarrow \rangle), \ \qquad |L_2 \rangle  = { 1\over \sqrt{2}} (| \uparrow \rangle - i | \downarrow \rangle). 
\end{equation}

The  quantum algebra specifies a relationship among  different axes, a property not explicitly addressed in our statistical model. 
In this system, the axes are independently uncertain, with no ``squeezing'' from states of one into another.
A textbook derivation from Eq. (\ref{paulispin})  gives independent   expressions for transverse displacement operators in a system in a definite  $\delta \hat X_3$ eigenstate for the two  state system considered here: 
\begin{equation}
 \langle \delta \hat X_1^2\rangle= l_P^2/2  \  \  {\rm and} \ \  \langle \delta \hat X_2^2\rangle=  l_P^2/2.
\end{equation}
If the symmetries of this quantum model apply in the real world, then the planar statistical model of this paper applies independently in two orthogonal planes, and  the  variance of classical twists around any axis
is given by Eq. (\ref{eqn:variance}), $\langle \delta  X_\perp^2\rangle = l_P^2/2$, as assumed in our model. The expected values of transverse operators also vanish to agree with the classical  inertial frame that is  used as the coordinate basis, so $\langle \delta  X_\perp\rangle = 0$ as assumed in the model.

The
serial preparation and measurement of the discrete quantum system creates a time series that approximates a world line  in a classical space-time.
In the discrete time series, each  system element is prepared in  an eigenstate of time, which is also the radial direction in space from the observer's world line. 
In an actual physical realization, the time  sequence does not remain in a superposition of transverse states;  because of decoherence, each light cone has to choose whether to turn left or right around each spatial axis at each step.
In the Copenhagen  version of measurement theory, this process  is described as  ``collapse'', but it can  also be regarded as a form of correlation, due to entanglement of the space-time with the observer's state.

The classical covariance (the spacing in time between the elements, or the  ``bin width'' in Eq. \ref{eqn:covariance}) follows from further appeal to the correspondence principle.  Each Planck step in proper time along a world line resets the state to the $|\uparrow\rangle$ or $|\downarrow\rangle$  eigenstate, which ensures that Planck scale causal symmetry of the light cone  states propagates to large scales. The variance in the null direction (the 3-axis of the algebra)  maps onto the same value in the classical time and space dimensions, with the standard relationship $c= l_P/ t_P$, which determines the discrete ``timestep'' $\Delta \mathbb{T}$ between light cones:
\begin{equation}
 \Delta \mathbb{T}^2= \langle \delta  \hat X_3^2\rangle/2c^2= t_P^2/4.
\end{equation}


Note that no step  of this construction has  invoked the  dynamical postulates of quantum mechanics.
The statistical correlations depend only on the more primitive quantum principles that govern Hilbert spaces of states and the operators that represent physical observables, as well as the  symmetries of the emergent space-time implicit in our choice of coordinate labels.  The physical analysis of the  discrete quantum system has  invoked superposition, correspondence, and measurement, but no explicit Hamiltonian. 
Time evolution as in the Schr\"odinger equation, and  the concepts of classical energy, momentum, dynamical conjugate pairings, angular momentum,  and  indeed $\hbar$   itself,  are all emergent or derived quantities.
Thus, the  calculation of the effects of these correlations on  states of general quantum fields (as opposed to the exactly null-propagating states whose properties depend only on causal structure, and are computed in our model) is highly nontrivial. 
The correspondence we have used to simplify this calculation  only works for one  observer world line at a time, since other observers are not perfectly inertial.  At the same time,   the construction of tangent light cones discussed above
 shows how a close correlation between nearby observers arises from  entangled light cone states.

\subsection{Time Asymmetry and Decoherence}\label{sec:timeasymmetry}

There is an apparent time asymmetry in the system we have described: the effect on a signal is not the same for outgoing and incoming parts of the same light path. The reason is simple to see: the path encounters light cones at different rates in the two directions. It is also apparent in the correlation functions, which show features at time intervals that do not correspond to the time intervals between reflections. 

The asymmetry can be traced to our basic hypothesis of covariance on future and not past light cones. Physically, this choice is motivated by the preparation of the system and matching to the cosmic inertial frame. The ``noise'' from the Planck scale corresponds to the new information arriving from the past about the rotational frame of the distant universe. A measurement is aligned with  the forward direction of time defined by quantum decoherence, which is also  the direction in which entropy increases. 
 The ``collapse of the wave function'' in this situation corresponds to matching with the rotation state of the rest of the universe--- the reconciliation of the observer's state with environmental information,  as shown in Fig. \ref{fig:twists}. 
It is important to note  that all of the measurable  predicted correlations, as well as all of the physical elements of the system,  are  symmetric  in time.


\subsection{Quantum Spookiness and Emergent Time}

It seems spooky, in the EPR sense,  to say that a space-time itself depends on an observer and a measurement path.
But in this sense, a perfectly classical, standard space-time also  represents a  spooky global conspiracy: the geometry of general relativity is a complete dynamical system that is independent of any observer.
Here,  a consistent picture emerges if we recognize that there must be   spooky (if still causally governed) entanglements among observers.
Our model of covariance formulates  a specific hypothesis for how macroscopic rotational relationships can emerge statistically  from the Planck scale, assuming only a covariance that respects causal relationships.
Nested causal surfaces of the entangled system resemble the intricately interlocked gears of a clockwork (as in Fig. \ref{twoslice}.)
Tangent displacements at causal boundaries  agree at every point for all of  the nested causal diamonds at that point. All observers will agree about physical observables, because all exotic ``motions,'' correlations and projections are defined in terms of radial light propagation and have no effect on radially propagating phase. This consistency also ensures that the whole system  closely resembles a continuous classical geometry on large scales. 
A path in space that begins and ends on an observer's  world line provides an operational statistical definition of an inertial frame,  built statistically out of relationships among quantum elements at the Planck scale. 

A well known feature of relational quantum geometry is that time itself is also part of the quantum system. This issue has been  avoided in the analysis above by analyzing the exotic motion in an observer's inertial frame, and covariance on Lorentz invariant light cones.   For a single world line, light phase along any radial direction represents a perfect ``light cone time'' clock. But for  two separate world lines, their clock phases only agree for light propagating along the radial separation between them.  There are differences in the transverse phase, which can be visualized as due an exotic transverse ``movement without motion'' of the underlying space at each world line relative to the other that causes a Planck scale random drift of transversely oriented light clocks.  In this sense, emergent time is not separate from  position, but also displays quantum weirdness\cite{Hogan:2015sra}.

To illustrate this point, consider a thought-experiment where the interference in  light that travels in the three arms of a triangle are converted into six real signals, which are later combined ``in software'' as Michelson or Sagnac modes. The exotic correlations should be the same as those computed here, since the correlation structure in our model depends only on classical causal structure. These measurements can be interpreted as comparisons between clocks in different places made in different directions, with exotic  spooky correlations caused by departures from inertial frames even of ``freely falling'' bodies.  As seen in the Sagnac and bent Michelson examples above, the total variance imprinted onto  a signal around a path  is an invariant that depends only on the shape of the path, but the detailed spectrum (the mapping onto an observer's local time) depends on how and where the signal is evaluated. 
We can say that the theory is Lorentz invariant, but the projections of states are observer dependent.
This thought experiment may some day become actualized in a spaceborne interferometer, although the exotic effect may be buried in gravitational wave foregrounds\cite{Hogan:2001jn}.

\subsection{Relation to  Emergent Gravity}


Our model is based on  an empty flat space-time, and does not include any  dynamics, matter, or curvature. It can be thought of as special relativity, quantized at the  Planck scale---  a covariant  quantum uncertainty of flat space.
On the other hand, because its statistical covariances are  built around a generally covariant classical causal structure--- that is, light cones and causal diamonds---   it provides a natural framework to define rotation, and predict exotic rotational correlations, in emergent  theories of gravity. 

It has been shown that  gravitational dynamics, as described by
the Einstein field equations, 
can be viewed as  thermodynamic relationships between entropy, temperature, and heat flux associated with    quantum elements at the Planck scale \cite{Jacobson1995,Verlinde2011,Padmanabhan:2013nxa}. 
 In  such a thermodynamic derivation of general relativity, the Einstein equations follow from the requirement that gravitational lensing curves space-time and distorts causal structure  by just the right amount  for the thermodynamic relations  to work.
Thermal excitations  appear at the macroscopic level as classical gravity, curvature and  acceleration. The thermodynamics do not depend on the detailed nature of the quantum elements. 

In this  view of emergent gravity,
our model could be a statistical description of quantum-geometrical fluctuations around a zero temperature ground state--- a  flat space-time with no matter.
The covariance on light cones that defines the measurement and collapse of a  space-time state by an observer is associated with  the same family of  null surfaces  that defines  thermodynamic quantities.  As in  thermodynamic gravity, our results do not depend on details of Planck scale quantum elements.  At the same time, our model already delves to a deeper level of detail, since it predicts new, observable quantum correlations from the Planck scale that do not exist in the classical theory and have not been predicted in thermodynamic gravity theories.  If we adopt  Jacobson's metaphor that general relativity resembles  a thermodynamic or hydrodynamic approximation, our model resembles  a theory of Brownian motion that reveals specific statistical signatures of atomic components.  The metaphor is not exact, because the system in our case is at zero temperature;  truly thermal  excitations would describe new effects of curvature. General relativity also includes effects of matter degrees of freedom, and their effects on causal structure, that are not included in our model.

In our model of emergent rotation, 
causal symmetry, Lorentz covariance, and the principle of equivalence are  built in--- states and displacements live on null surfaces---  but directionality and rotation are nonlocally and statistically defined.
Locality is in a sense also built into the model, but unlike classical relativity, it is inherited from an observer: distant space displays spooky, exotic, nonlocal spacelike correlations.
The principle of equivalence suggests that the model should approximately apply in any empty region of a curved space-time that is sufficiently small (compared to the curvature scale) to approximate a flat space-time, but still much larger than the Planck scale. For example, in a spinning Kerr black hole, the mean rotation on closed light path much smaller than the hole should almost agree with the spinning  ``local''   inertial frame of the classical metric, but should still display  almost the same exotic correlations and fluctuations around this mean value as the same path in flat space.  Of course, the same approximation applies very well to  interferometer measurements in terrestrial laboratories.

 \begin{acknowledgments}
We are grateful for much encouragement and challenging discussion from other members of the Holometer team. O.K. was supported by the Basic Science Research Program (Grant No. NRF-2016R1D1A1B03934333) of the National Research Foundation of Korea (NRF) funded by the Ministry of Education.  This work was supported by the John Templeton Foundation, and by the Department of Energy at Fermilab under Contract No. DE-AC02-07CH11359. 
\end{acknowledgments}

  \bibliographystyle{apsrev4-1}
\bibliography{lightcone}
\appendix

\newpage

\section{Detailed Analysis of the Bent Michelson Configuration\label{sec:bentspectra}}
The simple setup of the experiment design chosen for this example allows a detailed
set of analytical expressions for the integrals along each segment,
as explicitly shown in this appendix. 
Starting with Eqs. (\ref{eq:east-arm-theta}) and (\ref{eq:north-arm-theta}),
we see that the light propagation has a nonzero angular component
only along the ``bent'' portion of the east arm.  From this point on, we  drop the subscripts 1 and 2 for
the two arms, and only carry out calculations over the east arm (using
$\mathbf{x}_{1}(\tau)$ and $\Theta_{1}\left(\tau\right)$ above). So,
\begin{align}
C_{SS}(\tau_{0}\,|\,\ell_{P}) & =\int_{0}^{4T}d\tau'\,\frac{d\mathbb{T}}{d\tau'}(\tau',\,\mathbf{x}(\tau'))\,\Theta(\tau')\int_{\tau_{0}}^{4T+\tau_{0}}d\tau''\,\frac{d\mathbb{T}}{d\tau''}(\tau'',\,\mathbf{x}(\tau''-\tau_{0}))\,\Theta(\tau''-\tau_{0})\nonumber \\
 & \quad\times\Big\langle\frac{dX_{\perp}}{d\mathbb{T}}\big(\mathbb{T}(\tau',\,\mathbf{x}(\tau'))\big)\,\frac{dX_{\perp}}{d\mathbb{T}}\big(\mathbb{T}(\tau'',\,\mathbf{x}(\tau''-\tau_{0}))\big)\Big\rangle\label{eq:autocovar_bent}
\end{align}
where $\mathbf{x}(\tau)=\mathbf{x}_{1}(\tau)$ and $\Theta\left(\tau\right)=\Theta_{1}\left(\tau\right)$. 

From eq. \ref{eq:Ttau}, we derive inverse functions of $\mathbb{T}_{\shortrightarrow}(\tau'',\,\tau_{0})$
and $\mathbb{T}_{\shortleftarrow}(\tau'',\,\tau_{0})$, as follows:
\begin{align}
\tau'' & =\xi_{\shortrightarrow}(\mathbb{T},\,\tau_{0})\equiv\frac{1}{2}\left(\frac{T^{2}}{T+\tau_{0}-\mathbb{T}}+T+\tau_{0}+\mathbb{T}\right)\\
\tau'' & =\xi_{\shortleftarrow}(\mathbb{T},\,\tau_{0})\equiv\frac{1}{2}\left(\frac{T^{2}}{3T+\tau_{0}-\mathbb{T}}+3T+\tau_{0}+\mathbb{T}\right)
\end{align}
Again, similar equations hold for $\tau'$, with $\tau_{0}=0$. These
inverse functions do not hold over the entire light paths, but in
the ranges $T<\tau'<3T$ and $T<\tau''-\tau_{0}<3T$ where the geometrical
coupling is nonzero, they are well-defined.

We use the following notation for the projection factors:
\begin{equation}
\mathcal{P}_{\shortrightarrow}(\tau)\equiv\left[\frac{d\mathbb{T}_{\shortrightarrow}}{d\tau}(\tau,\,0)\right]\Theta_{\shortrightarrow}(\tau)\qquad\textrm{and}\qquad\mathcal{P}_{\shortleftarrow}(\tau)\equiv\left[\frac{d\mathbb{T}_{\shortleftarrow}}{d\tau}(\tau,\,0)\right]\Theta_{\shortleftarrow}(\tau)
\end{equation}
We now have  the tools to evaluate eq. \ref{eq:autocovar_bent},
the autocovariance of the signal for a layout that approximates an experiment now under construction.

Evaluating $C_{SS}(\tau_{0}\,|\,\ell_{P})$ is best done in segments.
We know $C_{SS}(\tau_{0}\,|\,\ell_{P})$ is symmetric, so we will
consider positive values of $\tau_{0}$ without any loss of generality.
Figure \ref{fig:layers} is a schematic representation of eq. \ref{eq:Ttau},
showing light cone slices of $\mathbb{T}$ versus the radial axis
$|\mathbf{x}(\tau)|$, where $\tau=\tau'$ and $\tau=\tau''-\tau_{0}$
for the two tracer photon trajectories separated by time lag $\tau_{0}$.
We see that $\mathbb{T}(\tau',\,\mathbf{x}(\tau'))$ or $\mathbb{T}(\tau'',\,\mathbf{x}(\tau''-\tau_{0}))-\tau_{0}$
stays at 0 until the bend mirror $(\tau=T)$, runs to $(2-\sqrt{2})T$
at the end mirror reflection $(\tau=2T$), and then to $2T$ at the
bend mirror on the way back $(\tau=3T)$. 

Following the covariance structure, the integrals should be performed
over the range of time where there are future light cones commonly
intersected by both tracer photon paths, within the constraints $T<\tau'<3T$
and $T<\tau''-\tau_{0}<3T$. These integrals should be subdivided
into separate segments whenever either tracer photon path flips from
outgoing to incoming, at $\tau=2T$. Converting the integrals into
intervals in $\mathbb{T}$, we see that it should be performed between
$\tau_{0}$ and $2T$, with segment divisions at $(2-\sqrt{2})T$
and $\tau_{0}+(2-\sqrt{2})T$. 

Correlations between segments in the same direction (outgoing or incoming)
give positive contributions to the integral, but anti-correlations
between outgoing and incoming segments give negative ones. This sharp
discontinuity in geometrical coupling at the end mirror reflection
gives rise to inflections in the autocorrelation whenever a segment
division is eliminated by increases in time lag $\tau_{0}$ (see Figure
\ref{fig:autospectra}). This is a feature not seen in the square
Sagnac spectra\textemdash{} although the mirror reflections happen
at roughly the same points, the geometrical coupling is continuous
there.

The functional support for $C_{SS}(\tau_{0}\,|\,\ell_{P})$ divides
into the following three ranges, making clear the origin of these
features:

\bigskip{}

\noindent \textbf{\large{}i)\enskip{}$0<\tau_{0}<(2-\sqrt{2})T$}\bigskip{}
\begin{align}
 & C_{SS}(\tau_{0}\,|\,\ell_{P})\nonumber \\
\nonumber \\
 & =\;\int_{\xi_{\shortrightarrow}(\tau_{0},\,0)}^{2T}\textrm{d}\tau'\,\mathcal{P}_{\shortrightarrow}(\tau')\int_{T+\tau_{0}}^{\xi_{\shortrightarrow}((2-\sqrt{2})T,\,\tau_{0})}\textrm{d}\tau''\,\mathcal{P}_{\shortrightarrow}(\tau''-\tau_{0})\,\times\,\left\langle \frac{dX_{\perp}}{d\mathbb{T}}\big(\mathbb{T}_{\shortrightarrow}(\tau',\,0)\big)\,\frac{dX_{\perp}}{d\mathbb{T}}\big(\mathbb{T}_{\shortrightarrow}(\tau'',\,\tau_{0})\big)\right\rangle \nonumber \\
 & \quad+\int_{2T}^{\xi_{\shortleftarrow}(\tau_{0}+(2-\sqrt{2})T,\,0)}\textrm{d}\tau'\,\mathcal{P}_{\shortleftarrow}(\tau')\int_{\xi_{\shortrightarrow}((2-\sqrt{2})T,\,\tau_{0})}^{2T+\tau_{0}}\textrm{d}\tau''\,\mathcal{P}_{\shortrightarrow}(\tau''-\tau_{0})\,\times\,\left\langle \frac{dX_{\perp}}{d\mathbb{T}}\big(\mathbb{T}_{\shortleftarrow}(\tau',\,0)\big)\,\frac{dX_{\perp}}{d\mathbb{T}}\big(\mathbb{T}_{\shortrightarrow}(\tau'',\,\tau_{0})\big)\right\rangle \nonumber \\
 & \quad+\int_{\xi_{\shortleftarrow}(\tau_{0}+(2-\sqrt{2})T,\,0)}^{3T}\textrm{d}\tau'\,\mathcal{P}_{\shortleftarrow}(\tau')\int_{2T+\tau_{0}}^{\xi_{\shortleftarrow}(2T,\,\tau_{0})}\textrm{d}\tau''\,\mathcal{P}_{\shortleftarrow}(\tau''-\tau_{0})\,\times\,\left\langle \frac{dX_{\perp}}{d\mathbb{T}}\big(\mathbb{T}_{\shortleftarrow}(\tau',\,0)\big)\,\frac{dX_{\perp}}{d\mathbb{T}}\big(\mathbb{T}_{\shortleftarrow}(\tau'',\,\tau_{0})\big)\right\rangle \qquad\quad\\
\nonumber \\
 & =\;c\,\ell_{P}\int_{\tau_{0}}^{(2-\sqrt{2})T}\Theta_{\shortrightarrow}(\xi_{\shortrightarrow}(\mathbb{T},\,0))\,\Theta_{\shortrightarrow}(\xi_{\shortrightarrow}(\mathbb{T},\,\tau_{0})-\tau_{0})\,d\mathbb{T}\nonumber \\
 & \quad+c\,\ell_{P}\int_{(2-\sqrt{2})T}^{\tau_{0}+(2-\sqrt{2})T}\Theta_{\shortleftarrow}(\xi_{\shortleftarrow}(\mathbb{T},\,0))\,\Theta_{\shortrightarrow}(\xi_{\shortrightarrow}(\mathbb{T},\,\tau_{0})-\tau_{0})\,d\mathbb{T}\nonumber \\
 & \quad+c\,\ell_{P}\int_{\tau_{0}+(2-\sqrt{2})T}^{2T}\Theta_{\shortleftarrow}(\xi_{\shortleftarrow}(\mathbb{T},\,0))\,\Theta_{\shortleftarrow}(\xi_{\shortleftarrow}(\mathbb{T},\,\tau_{0})-\tau_{0})\,d\mathbb{T}
\end{align}

\noindent \bigskip{}

\noindent \textbf{\large{}ii)\enskip{}}{\large{}$(2-\sqrt{2})T<\tau_{0}<\sqrt{2}T$}\bigskip{}
\begin{align}
 & C_{SS}(\tau_{0}\,|\,\ell_{P})\nonumber \\
\nonumber \\
 & =\;\int_{\xi_{\shortleftarrow}(\tau_{0},\,0)}^{\xi_{\shortleftarrow}(\tau_{0}+(2-\sqrt{2})T,\,0)}\textrm{d}\tau'\,\mathcal{P}_{\shortleftarrow}(\tau')\int_{T+\tau_{0}}^{2T+\tau_{0}}\textrm{d}\tau''\,\mathcal{P}_{\shortrightarrow}(\tau''-\tau_{0})\,\times\,\left\langle \frac{dX_{\perp}}{d\mathbb{T}}\big(\mathbb{T}_{\shortleftarrow}(\tau',\,0)\big)\,\frac{dX_{\perp}}{d\mathbb{T}}\big(\mathbb{T}_{\shortrightarrow}(\tau'',\,\tau_{0})\big)\right\rangle \nonumber \\
 & \quad+\int_{\xi_{\shortleftarrow}(\tau_{0}+(2-\sqrt{2})T,\,0)}^{3T}\textrm{d}\tau'\,\mathcal{P}_{\shortleftarrow}(\tau')\int_{2T+\tau_{0}}^{\xi_{\shortleftarrow}(2T,\,\tau_{0})}\textrm{d}\tau''\,\mathcal{P}_{\shortleftarrow}(\tau''-\tau_{0})\,\times\,\left\langle \frac{dX_{\perp}}{d\mathbb{T}}\big(\mathbb{T}_{\shortleftarrow}(\tau',\,0)\big)\,\frac{dX_{\perp}}{d\mathbb{T}}\big(\mathbb{T}_{\shortleftarrow}(\tau'',\,\tau_{0})\big)\right\rangle \qquad\quad\\
\nonumber \\
 & =\;c\,\ell_{P}\int_{\tau_{0}}^{\tau_{0}+(2-\sqrt{2})T}\Theta_{\shortleftarrow}(\xi_{\shortleftarrow}(\mathbb{T},\,0))\,\Theta_{\shortrightarrow}(\xi_{\shortrightarrow}(\mathbb{T},\,\tau_{0})-\tau_{0})\,d\mathbb{T}\nonumber \\
 & \quad+c\,\ell_{P}\int_{\tau_{0}+(2-\sqrt{2})T}^{2T}\Theta_{\shortleftarrow}(\xi_{\shortleftarrow}(\mathbb{T},\,0))\,\Theta_{\shortleftarrow}(\xi_{\shortleftarrow}(\mathbb{T},\,\tau_{0})-\tau_{0})\,d\mathbb{T}
\end{align}

\noindent \bigskip{}

\noindent \textbf{\large{}iii)\enskip{}}{\large{}$\sqrt{2}T<\tau_{0}<2T$}\bigskip{}
\begin{align}
 & C_{SS}(\tau_{0}\,|\,\ell_{P})\nonumber \\
\nonumber \\
 & =\;\int_{\xi_{\shortleftarrow}(\tau_{0},\,0)}^{3T}\textrm{d}\tau'\,\mathcal{P}_{\shortleftarrow}(\tau')\int_{T+\tau_{0}}^{\xi_{\shortleftarrow}(2T,\,\tau_{0})}\textrm{d}\tau''\,\mathcal{P}_{\shortrightarrow}(\tau''-\tau_{0})\,\times\,\left\langle \frac{dX_{\perp}}{d\mathbb{T}}\big(\mathbb{T}_{\shortleftarrow}(\tau',\,0)\big)\,\frac{dX_{\perp}}{d\mathbb{T}}\big(\mathbb{T}_{\shortrightarrow}(\tau'',\,\tau_{0})\big)\right\rangle \qquad\qquad\qquad\quad\;\,\\
\nonumber \\
 & =\;c\,\ell_{P}\int_{\tau_{0}}^{2T}\Theta_{\shortleftarrow}(\xi_{\shortleftarrow}(\mathbb{T},\,0))\,\Theta_{\shortrightarrow}(\xi_{\shortrightarrow}(\mathbb{T},\,\tau_{0})-\tau_{0})\,d\mathbb{T}
\end{align}
 
\end{document}